\documentclass[twocolumn,english,aps,prd,nofootinbib]{revtex4-1}
\usepackage{CJK}
\usepackage{amsfonts}
\usepackage{amsmath}
\usepackage{ dsfont }
\usepackage{hyperref}
\usepackage{amssymb}
\usepackage{xcolor}
\usepackage{xspace}
\usepackage{ragged2e}
\usepackage{relsize}
\usepackage[bottom]{footmisc}

\usepackage{tikz}
\usetikzlibrary{calc}
\newcommand{\nocontentsline}[3]{}
\newcommand{\tocless}[2]{\bgroup\let\addcontentsline=\nocontentsline#1{#2}\egroup}

\newcommand{\be}{\begin{equation}}
\newcommand{\ee}{\end{equation}}
\newcommand{\bea}{\begin{equation} \begin{aligned}}
\newcommand{\eea}{\end{aligned} \end{equation} }
\newcommand{\bi}{\begin{itemize}}
\newcommand{\ei}{\end{itemize}}

\renewcommand{\be}{\beta}
\newcommand{\al}{\alpha}
\newcommand{\bpm}{\begin{pmatrix}}
\newcommand{\epm}{\end{pmatrix}}
\newcommand{\eps}{\epsilon}

\renewcommand{\th}{\theta}
\newcommand{\lp}{\left(}
\newcommand{\rp}{\right)}

\newcommand{\del}{\partial}

\newcommand{\Tr}{\text{Tr} \ }

\newcommand{\mbf}[1]{\mathbf{#1}}

\usepackage{color}
\usepackage{graphicx}
\usepackage{verbatim}
\usepackage{amsmath}
\usepackage{amssymb}
\usepackage{wasysym}
\usepackage[caption=false]{subfig}
\usepackage{url}
\usepackage{bbold}
\usepackage{slashed}
\usepackage{epstopdf}
\usepackage{braket}
\usepackage{float}
\usepackage[percent]{overpic}

\DeclareRobustCommand{\App}[1]{App.~\ref{#1}}

\DeclareRobustCommand{\Fig}[1]{Fig.~\ref{#1}}

\DeclareRobustCommand{\Eq}[1]{Eq.~(\ref{#1})}

\DeclareRobustCommand{\Ref}[1]{Ref.~\cite{#1}}

\DeclareMathAlphabet\mathbfcal{OMS}{cmsy}{b}{n}

\RequirePackage[normalem]{ulem} 
\RequirePackage{color}\definecolor{RED}{rgb}{1,0,0}\definecolor{BLUE}{rgb}{0,0,1} 

\begin{document}

\title{Reentrant Correlated Insulators in Twisted Bilayer Graphene at 25T ($2\pi$ Flux)}

\author{Jonah Herzog-Arbeitman$^{1}$}
\author{Aaron Chew$^{1}$}
\author{Dmitri K. Efetov$^{2}$}
\author{B. Andrei Bernevig$^{1,3,4}$}

\affiliation{$^1$Department of Physics, Princeton University, Princeton, NJ 08544}
\affiliation{$^2$ ICFO - Institut de Ciencies Fotoniques, The Barcelona Institute of Science and Technology, Castelldefels, Barcelona 08860, Spain}

\affiliation{$^3$Department of Physics, Princeton University, Princeton, NJ 08544}
\affiliation{Donostia International Physics Center, P. Manuel de Lardizabal 4, 20018
Donostia-San Sebastian, Spain}
\affiliation{$^4$IKERBASQUE, Basque Foundation for Science, Bilbao, Spain}

\date{\today}

\begin{abstract}
Twisted bilayer graphene (TBG) is remarkable for its topological flat bands, which drive strongly-interacting physics at integer fillings, and its simple theoretical description facilitated by the Bistritzer-MacDonald Hamiltonian, a continuum model coupling two Dirac fermions. Due to the large moir\'e unit cell, TBG offers the unprecedented opportunity to observe reentrant Hofstadter phases in laboratory-strength magnetic fields near 25T. This Letter is devoted to magic angle TBG at $2\pi$ flux where the magnetic translation group commutes. We use a newly developed gauge-invariant formalism to determine the exact single-particle band structure and topology. We find that the characteristic TBG flat bands reemerge at $2\pi$ flux, but, due to the magnetic field breaking $C_{2z}\mathcal{T}$, they split and acquire Chern number $\pm 1$. We show that reentrant correlated insulating states appear at $2\pi$ flux driven by the Coulomb interaction at integer fillings, and we predict the characteristic Landau fans from their excitation spectrum. We conjecture that superconductivity can also be re-entrant at $2\pi$ flux.
\end{abstract}
\maketitle

\emph{Introduction.} Twisted bilayer graphene (TBG) is the prototypical moir\'e material obtained from rotating two graphene layers by an angle $\th$. Near the magic angle $\th = 1.05^\circ$, the two bands near charge neutrality flatten to a few meV, pushing the system into the strong-coupling regime and unravelling a rich landscape of correlated insulators and superconductors \cite{2018Natur.556...80C, Cao2018UnconventionalSI, Kim3364,2021NatPh..17..155K, balents2020superconductivity,liu2021orbital, Chu_2020}. Due to the large moir\'e unit cell, magnetic fluxes of $2\pi$ are achieved at only 25T.  In Hofstadter tight-binding models, such as the square lattice with Peierls substitution, the $2\pi$-flux and zero-flux models are equivalent, although the situation is more complicated in TBG \cite{PhysRevLett.125.236804}.  This begs the question: do insulating and superconducting phases of TBG repeat at 25T?

We study the Bistritzer-MacDonald (BM) Hamiltonian \cite{2011PNAS..10812233B}, describing the interlayer moir\'e-scale coupling of the graphene Dirac fermions within a single valley, which has established itself as a faithful model of the emergent TBG physics. We write the BM Hamiltonian in the particle-hole symmetric approximation as
\bea
\label{eq:HBM}
H_{BM}(\mbf{r}) = \bpm
-i \hbar v_F \pmb{\nabla} \cdot \pmb{\sigma} &  h.c. \\
\sum_{j=1}^3 T_j e^{2\pi i\mbf{q}_j \cdot \mbf{r}} & -i \hbar v_F \pmb{\nabla} \cdot \pmb{\sigma} \\
\epm \ .
\eea
Here $\mbf{q}_j = C_{3z}^{j-1} \mbf{q}_1$ are the inter-layer momentum hoppings, $\mbf{q}_1 =  (0, 4 \sin (\frac{\th}{2}) /3a_{g})$, and $a_{g} = .246$nm is the graphene lattice constant. The BM couplings  $T_1 = w_0 \sigma_0 + w_1 \sigma_1$, $T_{j+1} = \exp (\frac{2\pi i}{3} j \sigma_3) T_1\exp (-\frac{2\pi i}{3} j \sigma_3)$ act on the sublattice indices of the Dirac fermions, and $\sigma_j$ are the Pauli matrices. The lattice potential scale is $w_1 = 110$meV with $w_0/w_1 = .6$ - $.8$ \cite{PhysRevX.8.031087,2020arXiv200713390D} and the kinetic energy scale is $2\pi \hbar v_F |\mbf{q}_1| = 190$meV. The spectrum of $H_{BM}(\mbf{r})$ has been thoroughly investigated \cite{2018arXiv180710676S, PhysRevLett.122.106405, 2020arXiv200911301B,2020arXiv200911872S,2020arXiv201003589W,2020arXiv201107602C}.

The salient feature of the BM model from the Hofstadter perspective is the size of the moir\'e unit cell. After a unitary transform by $\text{diag}(e^{i \pi \mbf{q}_1 \cdot \mbf{r}}, e^{-i \pi \mbf{q}_1 \cdot \mbf{r}})$, $H_{BM}(\mbf{r})$ is put into Bloch form and is periodic under translations by $\mbf{a}_i$, the moir\'e lattice vectors \cite{PhysRevB.98.085435}. Near the magic angle, the moir\'e unit cell area $\Omega = |\mbf{a}_1 \times \mbf{a}_2|$ is a factor of $\th^{-2} \sim 3000$ times larger than the graphene unit cell. This dramatic increase in size brings the Hofstadter regime
\bea
\phi = e B \Omega/\hbar \sim 2\pi
\eea
within reach, showcasing physics which is only possible in strong flux \cite{PhysRevB.14.2239,PhysRevLett.125.236804,PhysRevLett.125.236805,PhysRevLett.86.147,andreibook,2020arXiv200613963L,moirehofexp}. Here $e/2\pi\hbar$ is the flux quantum (henceforth $e\! =\!\hbar\! =1\!$) and the magnetic field $B$ is near $25$T at $\phi = 2\pi$ and $\th = 1.05^\circ$. In the lattice Hofstadter problem, there is an \emph{exact} periodicity in flux depending on the orbitals \cite{PhysRevLett.125.236804}. This is no longer true in the continuum model \Eq{eq:HBM}. Nevertheless we find that the flat bands and correlated insulators are revived at $\phi = 2\pi$.

A constant magnetic field $\eps_{ij} \del_i A_j = B>0$ (repeated indices are summed) is incorporated into \Eq{eq:HBM} via the canonical substitution $-i \pmb{\nabla} \to \pmb{\pi} = -i \pmb{\nabla} - \mbf{A}(\mbf{r})$ yielding $H^\phi_{BM}$. Because the vector potential breaks translation symmetry, the spectrum in flux cannot be solved using Bloch's theorem. This problem has a long history with many approaches \cite{PhysRev.134.A1602,PhysRev.134.A1607,BROWN1969313,Streda_1982,1978PSSBR..88..757W,PhysRevB.76.115419,RevModPhys.82.1959,PhysRevB.52.14755,PhysRevB.84.035440,2019PhRvB.100c5115H,Crosse_2020,2021PhRvB.103L1405L}. However, we found that none accommodated the more demanding topological calculations essential for understanding TBG. Our separate work \Ref{secondpaper} contains technical calculations and proofs of formulae for the band structure, non-abelian Wilson loop, and many-body form factors. We apply the theory here to study the single-particle and many-body physics of TBG at $2\pi$ flux. Accompanying this paper, \Ref{exppaper} experimentally confirms our prediction of re-entrant correlated insulators in TBG at $2\pi$ flux. 

\emph{Magnetic Bloch Theorem.}
\begin{figure*}
\centering
\includegraphics[width=\textwidth]{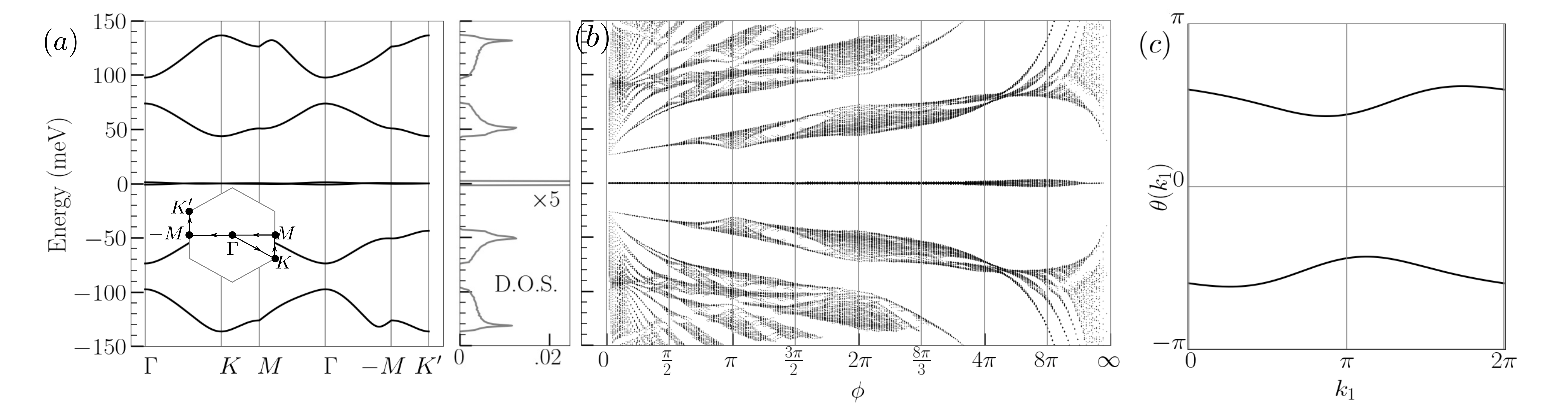}
\caption{TBG in flux. (a) The band structure and density of states at $\phi =2\pi$, $w_0/w_1 = 0.8$, and $\theta = 1.05^\circ$ reveal $\sim1.5$meV flat bands with a $40$meV gap. (b) The full Hofstadter spectrum shows the flat bands remain gapped at all flux. (c) Calculating the Wilson loop $W(k_1)$ of the two flat bands shows that, due to  $C_2\mathcal{T}$ breaking, the topology of the flat bands is trivial when they are connected.
}
\label{fig1}
\end{figure*}
In zero flux, the translation group of a crystal allows one to construct an orthonormal basis of momentum eigenstates labeled by $\mbf{k}$ in the Brillouin zone (BZ) and the spectrum is given by the Bloch Hamiltonian at each $\mbf{k}$. A similar construction can be followed at $2\pi$ flux where the magnetic translation group commutes. To begin, define the canonical momentum $\pi_\mu = - i \del_\mu - A_\mu$ and guiding centers $Q_\mu = \pi_\mu - B \eps_{\mu \nu} x_\nu$ which obey the (gauge-invariant) algebra
\bea
\label{eq:alg}
\null [\pi_\mu, \pi_\nu] = i B \eps_{\mu \nu}, \ [Q_\mu, Q_\nu] = -i B \eps_{\mu \nu}, \ [\pi_\mu, Q_\nu] = 0,
\eea
forming two decoupled algebras which are isomorphic to the free oscillator algebra. The kinetic term of \Eq{eq:HBM} contains only $\pi_\mu$ operators and commutes with the guiding centers $Q_\mu$. The Landau level ladder operators
\bea
a &= (\pi_x + i \pi_y)/\sqrt{2B}  , \quad a^\dag = (\pi_x - i \pi_y)/\sqrt{2B}
\eea
obeying $[a,a^\dag] = 1$ allow the Dirac Hamiltonian to be exactly solved in flux \cite{RevModPhys.82.1959}. Without a potential term, the $Q_\mu$ operators generate the macroscopic Landau level degeneracy. A potential term $U(\mbf{r})$ will break the degeneracy. If $U(\mbf{r})$ is periodic, the magnetic translation operators $T_{\mbf{a}_i} = e^{i \mbf{a}_i \cdot \mbf{Q}}$ commute with $H_{BM}^\phi$ because
\bea
T_{\mbf{a}_i} U(\mbf{r}) T^\dag_{\mbf{a}_i} = U(\mbf{r}+\mbf{a}_i) = U(\mbf{r}), \text{ and } \ [T_{\mbf{a}_i}, \pi_\mu] = 0
\eea
using \Eq{eq:alg} and the Baker-Campbell-Hausdorff (BCH) formula. The magnetic translation operators obey the projective representation
$T_{\mbf{a}_1} T_{\mbf{a}_2} = e^{i \phi} T_{\mbf{a}_2} T_{\mbf{a}_1}$ \cite{PhysRev.134.A1602}.
For generic flux, $T_{\mbf{a}_1}$ and $T_{\mbf{a}_2}$ do not commute, creating a characteristic fractal spectrum \cite{PhysRevB.14.2239}. Our interest in this work is the Hofstadter regime where $\phi = 2\pi$, the magnetic translation operators commute, and the spectrum consists of bands labeled by a ``momentum" $\mbf{k} = k_1 \mbf{b}_1 + k_2 \mbf{b}_2$, $k_i \in (-\pi,\pi)$ and $\mbf{a}_i\cdot\mbf{b}_j = \delta_{ij}$. To determine the band structure, one needs a basis of magnetic translation group irreps on infinite boundary conditions. Our results rest on the following construction at $\phi = 2\pi$:
\bea
\label{eq:MTGirreps}
\ket{\mbf{k}, n, \al, l} = \frac{1}{\sqrt{\mathcal{N}(\mbf{k})}} \sum_{\mbf{R}} e^{-i \mbf{k} \cdot \mbf{R}} T_{\mbf{a}_1}^{\mbf{R} \cdot \mbf{b}_1} T_{\mbf{a}_2}^{\mbf{R} \cdot \mbf{b}_2} \ket{n, \al, l}
\eea
where $\mbf{R}$ is the moir\'e Bravais lattice, $\al = A,B$ is the sublattice index, $l = \pm1$ is the layer index, and $n$ is the Landau level defined by $\ket{n,\al,l} = \frac{a^{\dag \,n}}{\sqrt{n!}} \ket{0,\al,l}, \ a\ket{0,\al,l} = 0$. A similar construction was used in \Ref{PhysRevLett.125.236804} to identify a projective representation of the magnetic space group $1'$ in the Hofstadter Hamiltonian of a tight-binding model. The states in \Eq{eq:MTGirreps} are magnetic translation group eigenstates obeying $T_{\mbf{a}_i} \ket{\mbf{k},n ,\al, l} = e^{i \mbf{k} \cdot \mbf{a}_i} \ket{\mbf{k},n,\al, l}$, which immediately proves their orthogonality at different $\mbf{k}$. Orthogonality at different $n$ follows because $\ket{\mbf{k}, n}$ are eigenstates of the Hermitian operator $a^\dag a$ with eigenvalue $n$. The normalization $\mathcal{N}(\mbf{k})$ is determined by requiring orthonormality $\braket{\mbf{k}', m|\mbf{k}, n} = (2\pi)^2 \delta_{mn} \delta(\mbf{k} - \mbf{k}')$ and can be expressed in terms of theta functions (\App{app:formula}). We find that $\mathcal{N}(\mbf{k}^*) = 0$, indicating that the states are not well-defined at $\mbf{k}^* = \pi \mbf{b}_1+\pi \mbf{b}_2$. This is because the states in \Eq{eq:MTGirreps} are built from Landau levels $\ket{n, \alpha, l}$ which carry a Chern number, but Chern states cannot be periodic and well-defined everywhere in the BZ \cite{Brouder_2007}. On infinite boundary conditions, the point $\mbf{k}^*$ is a set of measure zero in the BZ, and we prove in \Ref{secondpaper} that the basis in \Eq{eq:MTGirreps} is complete with the exception of pathological examples that do not occur when the wavefunctions are suitably smooth.

The basis states in \Eq{eq:MTGirreps} yield a simple expression for the magnetic Bloch Hamiltonian
\bea
\label{eq:Ham2pi}
\null (2\pi)^2 \delta(0) [H_{BM}^{\phi = 2\pi}(\mbf{k})]_{mn, \al\be, ll'} = \braket{\mbf{k},m,
\al,l| H^{\phi=2\pi}_{BM}| \mbf{k},n,\be,l'} \ .
\eea
The matrix elements of \Eq{eq:Ham2pi} can be computed exactly to obtain an expression for $H_{BM}^{\phi=2\pi}(\mbf{k})$ (\App{app:formula}). Truncating to $N_{LL}$ Landau levels, we obtain a finite $4 N_{LL} \times 4 N_{LL}$ matrix that can be diagonalized at each $\mbf{k}$ to produce a band structure. This is similar to the zero flux expansion of $H_{BM}$ on a plane wave basis, where high momentum modes are truncated. As computed in \Fig{fig1}a, the famous flat bands of magic-angle TBG remain at $2\pi$ flux suggesting that the system will be dominated by strong interactions. We use the open momentum space technique \cite{2021PhRvB.103L1405L} to obtain the Hofstadter spectrum (\Fig{fig1}b) which shows the evolution of the higher energy passive bands. At $2\pi$ flux, full density Bloch-like flat bands reappear at charge neutrality and are the focus of this work.

\emph{Topology of the Flat bands.} Similar to the zero flux TBG flat bands, the reentrant flat bands at $2\pi$ flux have a very small bandwidth of $\sim1$ meV. However, their topology is quite different due to the breaking of crystalline symmetries by magnetic field. Let us review the zero flux model. \Ref{2018arXiv180710676S} showed that the space group $p6'2'2$ of the BM Hamiltonian (\Eq{eq:HBM}) was generated by $C_{3z}, C_{2x},$ and $C_{2z}\mathcal{T}$ and also featured an approximate unitary particle-hole operator $P$. Notably, $C_{2z}\mathcal{T}$ alone is sufficient to protect the gapless Dirac points and fragile topology of the flat bands \cite{2018arXiv180710676S}.

Because a perpendicular magnetic field is reversed by time-reversal and $C_{2x}$ symmetries (while it is invariant under in-plane rotations), the $C_{2x}$ and $C_{2z} \mathcal{T}$ symmetries are broken in flux \cite{PhysRevLett.125.236804}. Thus, the space group of $H_{BM}^\phi$ is reduced to $p31m'$ which is generated by $C_{3z}$ and $M\mathcal{T} \equiv C_{2x}C_{2z} \mathcal{T}$. $P$ also remains a symmetry. Without $C_{2z}\mathcal{T}$, the system changes substantially. The most direct way to assess the topology at $2\pi$ flux is to calculate the non-Abelian Wilson loop. To do so, we need an expression for the Berry connection $\mathcal{A}^{MN}(\mbf{k})$ where $M,N$ index the occupied bands. At $2\pi$ flux, the Berry connection ${\cal A}_i = {\mathbf b}_i \cdot {\cal A}$ contains new contributions \cite{secondpaper}:
\bea
\label{eq:berrycon}
\mathcal{A}_i^{MN}(\mbf{k}) &= [U^\dag(\mbf{k}) (i \del_{k_i} - \eps_{ij} \tilde{Z}_j) U(\mbf{k})]^{MN} \\
& \qquad\qquad - \delta^{MN} \eps_{ij}   \del_{k_j} \log \sqrt{\mathcal{N}(\mbf{k})}
\eea
where $U(\mbf{k})$ is the matrix of eigenvectors and $M,N$ span the occupied bands. In the case of the TBG flat bands, $U(\mbf{k})$ is a $4 N_{LL} \times 2$ matrix. The Abelian term in the second line of \Eq{eq:berrycon} is an exact expression for the Berry connection of a Landau level which is discussed at length in \Ref{secondpaper} and accounts for the Chern number of the basis states. The non-Abelian term $\tilde{Z}_j$ acts nontrivially on the Landau level indices (\App{app:formula}). We numerically calculate the Wilson loop \cite{Alexandradinata:2012sp} over the flat bands in \Fig{fig1}(c) which shows no winding. Hence the fragile topology of the flat bands, which was protected by $C_{2z} \mathcal{T}$, is broken in flux. However, we calculate that the neighboring passive bands are gapped (unlike at zero flux) and carry nonzero Chern numbers (\App{app:moreplots}). They are dispersive Landau levels originating from the Rashba point of the passive bands at zero flux \cite{2020arXiv200713390D}.

To gain a deeper understanding of the topology at $2\pi$ flux, we study the band representation $\mathcal{B}$ with topological quantum chemistry \cite{2017Natur.547..298B,Aroyo:firstpaper,Aroyo:xo5013}. First, \Fig{fig1}b demonstrates that the flat bands remain gapped from all other bands in flux. This is despite the fragile topology of TBG, verifying the prediction of \Ref{PhysRevLett.125.236804}. $C_{2z}$ symmetry, however is sufficient to protect a gap closing in concert with the fragile topology. Thus $\mathcal{B}$ can be simply obtained by reducing the band representation of TBG in zero flux derived in \Ref{2018arXiv180710676S} to $p31m'$. We find
\bea
\label{eq:BR}
\mathcal{B} = 2 \Gamma_1 + K_2 + K_3 + K_2' + K_3' = A_{2b} \uparrow p31m'
\eea
which is an elementary band representation and is not topological. The irreps are defined
\bea
\begin{array}{r|rr}
3m'&1& C_{3z} \\
\hline
\Gamma_1 &1&1\\
\end{array}, \quad \begin{array}{r|rr}
3 &1& C_{3z} \\
\hline
K_2 &1 & e^{\frac{2\pi i}{3}}\\
K_3 &1 & e^{-\frac{2\pi i}{3}}\\
\end{array},\quad \begin{array}{r|rr}
3 &1& C_{3z} \\
\hline
K'_3 &1 & e^{\frac{2\pi i}{3}}\\
K'_2 &1 & e^{-\frac{2\pi i}{3}}\\
\end{array}
\eea
and $A_{2b}$ denotes two one-dimensional irreps of $s$ orbitals placed at the corners of the moir\'e unit cell, which matches the charge distribution at zero flux \cite{2018arXiv180710676S,2018PhRvX...8c1088K,PhysRevX.8.031087}. Another simple observation is that the total Chern number of the two flat bands is zero, so the flat bands cannot be modeled by decoupled Landau levels despite the strong flux, which demonstrates the importance of our exact approach. Consulting the \href{https://www.cryst.ehu.es/cgi-bin/cryst/programs/mbandrep.pl}{Bilbao Crystallographic Server}, we observe that \Eq{eq:BR} is decomposable in momentum space \cite{2018PhRvL.120z6401C,2018PhRvB..97c5139C,Bradlyn_2019}, meaning that $\mathcal{B}$ may be split into two disconnected bands:
\bea
\label{eq:Bsplit}
\mathcal{B} = \mathcal{B}_+ + \mathcal{B}_- = (\Gamma_1 \!+\! K_2 \!+\! K_3') + (\Gamma_1 \!+\! K_3 \!+\! K_2')
\eea
where $\mathcal{B}_\pm$ carries Chern number $C = \pm 1$ mod $3$ \cite{2012PhRvB..86k5112F}. The irreps of $\mathcal{B}_\pm$ at the $K$ and $K'$ points are related by the anti-unitary operator $M\mathcal{T}$ which obeys $C_{3z} M\mathcal{T} = M\mathcal{T} C_{3z}^\dag$, so \Eq{eq:Bsplit} is the only allowed decomposition. We show below that the addition of $P$, which is not part of the irrep classification (it is not a crystallographic symmetry), forbids this splitting.

\Eq{eq:Bsplit} suggests a remarkable similarity to the topology of the flat bands at zero flux, where $C_{2z}\mathcal{T}$ enforces \emph{connected} bands whose Wilson loop eigenvalues wind in opposite directions \cite{PhysRevB.99.155415,2018arXiv180710676S,2020arXiv200911872S,2019PhRvX...9b1013A}. $C_{2z}\mathcal{T}$ is crucial to protecting the fragile topology, which would otherwise be trivialized from the cancelation of the winding. At $2\pi$ flux, breaking $C_{2z}\mathcal{T}$ destroys the fragile topology but allows the bands to split and carry opposite non-zero Chern numbers. Thus in flux, the fragile topology in the two TBG flat bands is replaced by stable topology as the bands split and acquire a Chern number. These bands carry \emph{opposite} Chern number, but they cannot annihilate with each other: $M\mathcal{T}$ symmetry ensures any band touching come in pairs so the Chern numbers can only change in multiples of two. To understand the mechanism which splits the flat bands, we re-examine $P$ which has so far been neglected. $P$ is not an exact (but still a very good) symmetry of TBG and only anti-commutes when terms of $O(\th)$ are dropped \cite{2018arXiv180710676S,2020arXiv200911872S}. We incorporate the exact $\th$ dependence into the kinetic terms of \Eq{eq:HBM}, breaking $P$ and opening a $\sim.5$meV gap between the flat bands at $K$ and $K'$ and verify the Chern number decomposition in \Eq{eq:Bsplit} from the Wilson loop (\App{app:PH}).

The particle-hole approximation prevents the Chern decomposition because $P$ and $C_{3z}$ enforce gapless points at $K$ and $K'$ as we now show. Observe that  the $K$ and $K'$ points are symmetric under the anti-commuting symmetry $\mathcal{P} = P M\mathcal{T}$  because $P$ takes $\mbf{k} \to - \mbf{k}$ and  $M\mathcal{T}$ takes $(k_x,k_y) \to (k_x,- k_y)$ \cite{2018arXiv180710676S}. $\mathcal{P}$ is anti-unitary and obeys $C_{3z} \mathcal{P} = \mathcal{P} C_{3z}^\dag$. As such, a state $\ket{\omega}$ of energy $E\neq0$ and $C_{3z}$ eigenvalue $\omega$ ensures a distinct state $\mathcal{P} \ket{\omega}$ with $C_{3z}$ eigenvalue $\omega$ and energy $-E$. Thus all states at $E\neq 0$ come in $\mathcal{P}$-related pairs with the same $C_{3z}$ eigenvalue. We see that the irreps of $\mathcal{B}$ at $K$ and $K'$ cannot be gapped (they are pinned to $E=0$) without violating $\mathcal{P}$ because they have different $C_{3z}$ eigenvalues.

\emph{Coulomb Groundstates.} We have derived the spectrum and topology of TBG at $2\pi$ flux, thoroughly studying its single-particle physics. When considering many-body states, we must include the spin and valley degrees of freedom. The low energy states in TBG come from the two graphene valleys which we index by $\eta = \pm1$. The valleys are interchanged by $C_{2z}$ which is unbroken by flux, and hence the flat bands are each four-fold degenerate. To split the degeneracy, we consider adding the interaction
\bea
H_{int} = \frac{1}{2\Omega_{tot}} \sum_{\mbf{q}} V(\mbf{q}) \bar\rho_{-\mbf{q}} \bar\rho_{\mbf{q}}, \quad \bar\rho_{\mbf{q}}=  \int d^2r \, e^{-i \mbf{q} \cdot \mbf{r}} \bar n(\mbf{r})
\eea
where $V(\mbf{q}) >0$ is the screened Coulomb potential \cite{2020arXiv200912376B,2021Sci...371.1261L}, $\bar{n}(\mbf{r})$ is the total electron density (summed over valley and spin) measured from charge neutrality, and $\Omega_{tot}$ is the area of the sample. We now discuss the symmetries of the many-body Hamiltonian. In zero flux, the single-particle and interaction terms conserve spin, charge, and valley, so there is an exact $U(2) \times U(2)$ symmetry group. It is also natural to work in a strong coupling expansion where we project $H_{int}$ onto the two flat bands and neglect their kinetic energy entirely. This is a very reliable approximation because the bandwidth is $O(1)$ meV and the interaction strength is $\sim 20$meV. In this limit, $C_{2z} P$ commutes with the projected $H_{int}$ operator and the symmetry group is promoted to $U(4)$\cite{2020arXiv200912376B,2020arXiv200909413V}.

We now discuss the fate of the $U(4)$ symmetry in flux. At $B \sim 25$T, the Zeeman effect shifts the energy of the spin $\pm1/2$ electrons by $\pm \mu_B B = \pm 1.4$meV where $\mu_B$ is the Bohr magneton. This shift is comparable to the bandwidth, so it is consistent to neglect both at leading order.  (The Zeeman term will choose the spin-polarized states out of the $U(4)$ manifold.) Similarly, although $P$-breaking terms allow the flat bands to gap at $2\pi$ flux, the kinetic energy remains $\leq 3$meV, so it is consistent to neglect the single-particle Hamiltonian (including particle-hole breaking terms) as a first approximation. The last effect to address is twist angle homogeneity which has recently come under scrutiny \cite{PhysRevResearch.2.023325,2020arXiv201209885P,2020arXiv200502406P}. Experiments indicate that even in high quality devices, the moir\'e twist angle $\th$ varies locally up to $.1^\circ$ \cite{2020Natur.581...47U,2020arXiv200809761K,PhysRevResearch.3.013153}, varying the magnetic field at $\phi =2\pi$ between $25-30$T for $\th \in (1^\circ,1.1^\circ)$. In a realistic sample with domains of varying $\th$ at constant $B$, it is reasonable to expect non-ideal flat bands with higher bandwidth. However, the large interaction strength and gap to the passive bands still makes the strong coupling expansion appropriate. In this limit, the $U(4)$ symmetry is intact.

An analytic study of the strong-coupling problem is possible because $H_{int}$ is positive semi-definite \cite{PhysRevLett.122.246401}. Following \Ref{2020arXiv200914200B}, we will study exact eigenstates at fillings $\nu = 0, + 2,+4$ (the $-\nu$ states follows from many-body particle-hole symmetry \cite{2020arXiv200912376B}) and derive the excitation spectrum there --- effectively determining the complete renormalization of band structure by the Coulomb interaction. \Ref{2020arXiv200914200B} was also able to study odd integer fillings perturbatively using the chiral symmetry at $w_0 = 0$ \cite{PhysRevLett.122.106405,2020arXiv201003589W,2020arXiv200911301B,2021arXiv210610650S}. The chiral limit $w_0 = 0$ at $2\pi$ flux is topologically distinct from the physical regime $w_0/w_1 = .6$ - $.8$ (unlike at zero flux) so this approach is inapplicable \cite{2021arXiv210610650S}. We leave the odd fillings to future work.

The many-body calculation at $2\pi$ flux is tractable using a gauge-invariant expression for $H_{int}$ and the form factors. Following \Ref{2020arXiv200913530L}, we produce exact many-body insulator eigenstates of the projected Coulomb Hamiltonian at filling $\nu \in (-4,4)$:
\bea
\label{eq:exacteigstate}
\ket{\Psi_{\nu}} &= \prod_{\mbf{k}}  \prod_{j}^{(4+\nu)/2}  \gamma^\dag_{\mbf{k},+, \eta_j,s_j}  \gamma^\dag_{\mbf{k},-, \eta_j,s_j}  \ket{0}
\eea
where the electron operators $\gamma^\dag_{\mbf{k}, M, \eta,s}$ create a state at momentum $\mbf{k}$, valley $\eta$, and spin $s$ in the $M = \pm1$ band. The states $\ket{\Psi_{\nu}}$ fully occupy the two flat bands for arbitrary $\eta_j,s_j$ forming a $U(4)$ multiplet. Including valley and spin, there are $8$ flat bands; state $\ket{\Psi_\nu}$ fills $(4 + \nu)/2$ of them. At $\nu = 0$, $\ket{\Psi_{0}}$ must be a groundstate because $H_{int}$ is positive semi-definite and $H_{int}\ket{\Psi_{0}}  = 0$. At $\nu = \pm4$ where the system is a band insulator, $\ket{\Psi_{\pm4}}$ are trivially groundstates because they are completely empty/occupied respectively. The $\ket{\Psi_{\pm2}}$ states are exact eigenstates, and we argue they are groundstates using the flat metric condition (FMC) \cite{2020arXiv200913530L} which assumes the Hartree potential of the flat bands is trivial. \Ref{2020arXiv200912376B} found that the FMC holds reliably at zero flux, and we check that the FMC is similarly reliable at $2\pi$ flux \cite{secondpaper}.

The exact eigenstates $\ket{\Psi_\nu}$ enable us to compute the excitation spectrum near filling $\nu$. The Hamiltonian $R^\eta_+(\mbf{k})$ governing the $+1$ charge spectrum is defined
\bea
\label{eq:charge1}
\null [H_{int} - \mu N, \gamma^\dag_{\mbf{k},M, s, \eta}] \ket{\Psi_\nu} &\equiv \frac{1}{2} \sum_N  \gamma^\dag_{\mbf{k},N, s, \eta} [R^\eta_+(\mbf{k})]_{NM} \ket{\Psi_\nu}  \\
\eea
where $\eta, s$ are \emph{unoccupied} indices in $\ket{\Psi_\nu}$ and $\mu$ is the chemical potential (\App{app:BMchargeR}). Counting the flavors in \Eq{eq:exacteigstate}, at filling $\nu$ the charge $\pm1$ excitations come in multiples of $(4\mp\nu)/2$. We give an explicit expression for $R^\eta_{\pm}(\mbf{k})$, the $\pm1$ charge excitation Hamiltonian, in \App{app:BMchargeR}.

\begin{figure}
\centering
\begin{overpic}[height=0.25\textwidth]{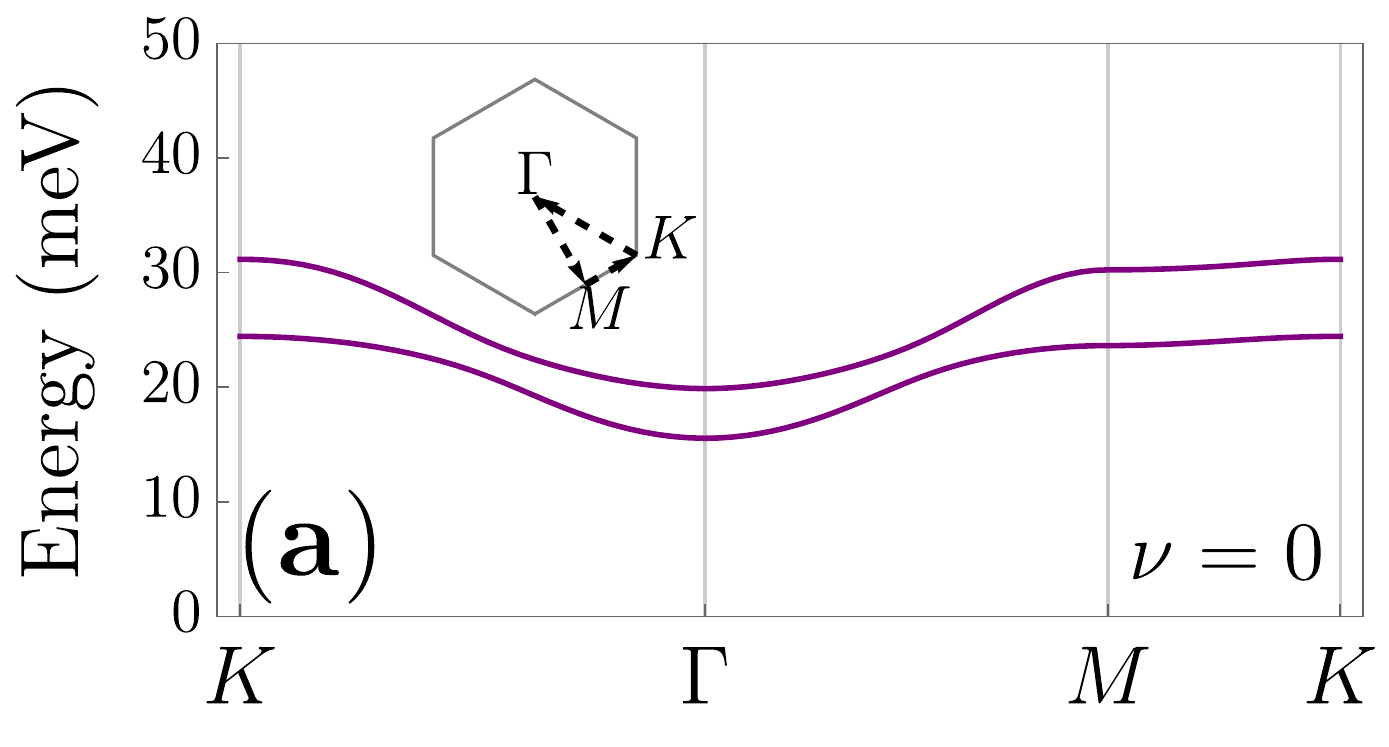}
\end{overpic}
\begin{overpic}[height=0.213\textwidth]{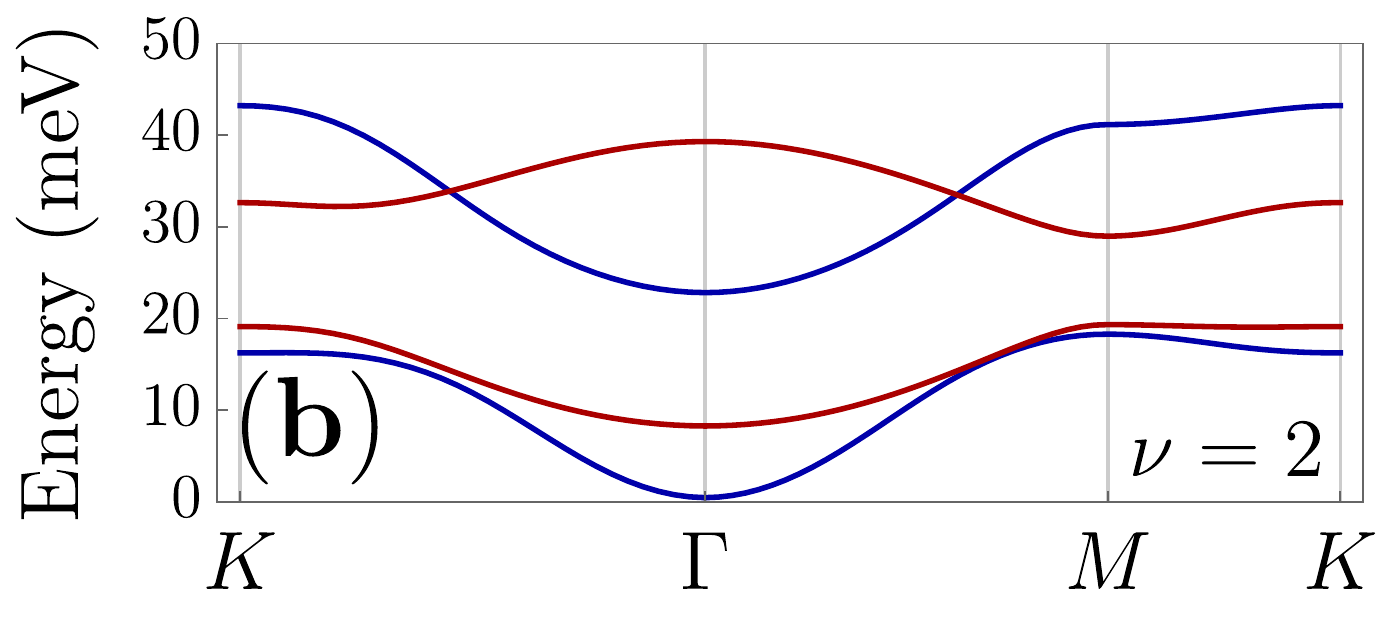}
\end{overpic}
\caption{$\frac{1}{2}R^\eta_\pm(\mbf{k})$ spectra at $w_0/w_1=.71$: positive energies denote a charge gap. (a) At $\nu = 0$, the charge $\pm 1$ excitations are identical and feature a massive particle dispersion at the $\Gamma$ point. Degeneracies are lifted because flux breaks $C_{2z}\mathcal{T}$. (b) At $\nu = 2$, the charge $-1$ excitation (red) has a large mass, strongly suppressing the Landau fans pointing towards charge neutrality, while the $+1$ excitation (blue) is lighter by a factor of 3 with a mass of $\sim 200$meV in units where $v_F =1$. The $+1$ charge gap at $\nu=2$ is $\sim .5$meV or roughly $5$K. 
}
\label{fig2}
\end{figure}

The excitation spectra in \Fig{fig2} describe the behavior of TBG at densities close to $\nu$, giving distinctive signatures in the Landau fans emanating from the $\ket{\Psi_\nu}$ insulators \cite{2020arXiv200713390D,1978PSSBR..88..757W,2018arXiv181111786L}. At $\nu = 0$, the $\pm1$ charge excitations are identical and their dispersion features a charge gap to a band with a quadratic minima at the $\Gamma$ point. Hence at low densities, there are $(4\mp0)/2 = 2$ massive quasi-particles, counting the degenerate charge excitations in different spin-valley flavors. As the flux is increased, the massive quadratic excitations form Landau levels (quantum Hall states), leading to Landau fans away from $\nu=0$ in multiples of 2 --- half the Landau level degeneracy of TBG near $B=0$. The gap between the two excitation bands at $\Gamma$ depends on $w_0/w_1$. \Fig{fig2}a shows the generic case at $w_0/w_1 = .71$, but at $w_0/w_1 = .8$ the two bands are nearly degenerate at $\Gamma$ (\App{app:BMchargeR}). At $\nu = 2$, the $-1$ excitation (towards charge neutrality) has a large mass which reduces the gap between Landau levels and masks would-be insulating states. However, the $+1$ excitation has a smaller effective mass and will create Landau levels in multiples of $(4-2)/2 = 1$. We do not discuss excitations above $\nu=4$ here because they fill the passive bands, and we check that the charge $-1$ excitation below $\nu=4$ (not shown) is gapped with a very large mass. We note that, with $C_{2z}\mathcal{T}$ at zero flux, the excitation bands must be degenerate at the $\Gamma$ point \cite{2020arXiv200914200B,2021arXiv210401145K}. This is not the case at $2\pi$ flux where $C_{2z}\mathcal{T}$ is broken. Based on the $U(4)$ symmetry which determines the $(4\mp \nu)/2$ degeneracy of the excitations, the breaking of $C_{2z}\mathcal{T}$ which allows the bands to be gapped at $\Gamma$, and the large mass of excitations towards charge neutrality, we predict the Landau fans emerging from $\nu = 0$ and $\nu=2$ away from charge neutrality to have degeneracies $2$ and $1$ respectively, half that of TBG. Comparing with the zero-flux charge excitations in \Ref{2020arXiv200914200B}, we find that the effective masses of the excitations are larger by a factor of $\sim 2$ at $2\pi$ flux, making the Landau fans more susceptible to disorder.  

\emph{Discussion.} We used an exact method to study TBG at $2\pi$ flux, yielding comprehensive results for the single-particle and many-body physics. Recently, interest in reentrant superconductivity and correlated phases in strong flux has invigorated research in moir\'e materials \cite{2021arXiv210501243C,2021arXiv210312083C}. Our formalism makes it possible to study such phenomena with the tools of modern band theory and without recourse to approximate models. We find that the emblematic topological flat bands and correlated insulators of TBG are re-entrant at $\phi = 2\pi$, providing strong evidence that magic angle physics recurs at $\sim 25$T. This leads us to conjecture that superconductivity, which occurs at $\phi = 0$ upon doping correlated insulating states, may also be reentrant at $2\pi$ flux, as discussed in \Ref{exppaper}.

\emph{Acknowledgements.} We thank Zhi-Da Song for early insight and Luis Elcoro for useful discussions. B.A.B. and A.C. were supported by the ONR Grant No. N00014-20-1-2303, DOE Grant No. DESC0016239, the Schmidt Fund for Innovative Research, Simons Investigator Grant No. 404513, the Packard Foundation, the Gordon and Betty Moore Foundation through Grant No. GBMF8685 towards the Princeton theory program, and a Guggenheim Fellowship from the John Simon Guggenheim Memorial Foundation. Further support was provided by the NSF-MRSEC Grant No. DMR-1420541 and DMR-2011750, BSF Israel US foundation Grant No. 2018226, and the Princeton Global Network Funds. JHA is supported by a Marshall Scholarship funded by the Marshall Aid Commemoration Commission.

\let\oldaddcontentsline\addcontentsline
\renewcommand{\addcontentsline}[3]{}
\bibliography{finalbib}

\begin{thebibliography}{67}
\providecommand{\natexlab}[1]{#1}
\providecommand{\url}[1]{\texttt{#1}}
\expandafter\ifx\csname urlstyle\endcsname\relax
  \providecommand{\doi}[1]{doi: #1}\else
  \providecommand{\doi}{doi: \begingroup \urlstyle{rm}\Url}\fi

\bibitem[{Cao} et~al.(2018){Cao}, {Fatemi}, {Demir}, {Fang}, {Tomarken}, {Luo},
  {Sanchez-Yamagishi}, {Watanabe}, {Taniguchi}, {Kaxiras}, {Ashoori}, and
  {Jarillo-Herrero}]{2018Natur.556...80C}
Yuan {Cao}, Valla {Fatemi}, Ahmet {Demir}, Shiang {Fang}, Spencer~L.
  {Tomarken}, Jason~Y. {Luo}, Javier~D. {Sanchez-Yamagishi}, Kenji {Watanabe},
  Takashi {Taniguchi}, Efthimios {Kaxiras}, Ray~C. {Ashoori}, and Pablo
  {Jarillo-Herrero}.
\newblock {Correlated insulator behaviour at half-filling in magic-angle
  graphene superlattices}.
\newblock \emph{\nat}, 556\penalty0 (7699):\penalty0 80--84, Apr 2018.
\newblock \doi{10.1038/nature26154}.

\bibitem[Cao et~al.(2018)Cao, Fatemi, Fang, Watanabe, Taniguchi, Kaxiras, and
  Jarillo-Herrero]{Cao2018UnconventionalSI}
Yuan Cao, V.~Fatemi, S.~Fang, K.~Watanabe, T.~Taniguchi, E.~Kaxiras, and
  P.~Jarillo-Herrero.
\newblock Unconventional superconductivity in magic-angle graphene
  superlattices.
\newblock \emph{Nature}, 556:\penalty0 43--50, 2018.

\bibitem[Kim et~al.(2017)Kim, DaSilva, Huang, Fallahazad, Larentis, Taniguchi,
  Watanabe, LeRoy, MacDonald, and Tutuc]{Kim3364}
Kyounghwan Kim, Ashley DaSilva, Shengqiang Huang, Babak Fallahazad, Stefano
  Larentis, Takashi Taniguchi, Kenji Watanabe, Brian~J. LeRoy, Allan~H.
  MacDonald, and Emanuel Tutuc.
\newblock Tunable moir{\'e} bands and strong correlations in small-twist-angle
  bilayer graphene.
\newblock \emph{Proceedings of the National Academy of Sciences}, 114\penalty0
  (13):\penalty0 3364--3369, 2017.
\newblock ISSN 0027-8424.
\newblock \doi{10.1073/pnas.1620140114}.
\newblock URL \url{https://www.pnas.org/content/114/13/3364}.

\bibitem[{Kennes} et~al.(2021){Kennes}, {Claassen}, {Xian}, {Georges},
  {Millis}, {Hone}, {Dean}, {Basov}, {Pasupathy}, and
  {Rubio}]{2021NatPh..17..155K}
Dante~M. {Kennes}, Martin {Claassen}, Lede {Xian}, Antoine {Georges}, Andrew~J.
  {Millis}, James {Hone}, Cory~R. {Dean}, D.~N. {Basov}, Abhay~N. {Pasupathy},
  and Angel {Rubio}.
\newblock {Moir{\'e} heterostructures as a condensed-matter quantum simulator}.
\newblock \emph{Nature Physics}, 17\penalty0 (2):\penalty0 155--163, January
  2021.
\newblock \doi{10.1038/s41567-020-01154-3}.

\bibitem[Balents et~al.(2020)Balents, Dean, Efetov, and
  Young]{balents2020superconductivity}
Leon Balents, Cory~R Dean, Dmitri~K Efetov, and Andrea~F Young.
\newblock Superconductivity and strong correlations in moir{\'e} flat bands.
\newblock \emph{Nature Physics}, 16\penalty0 (7):\penalty0 725--733, 2020.

\bibitem[Liu and Dai(2021)]{liu2021orbital}
Jianpeng Liu and Xi~Dai.
\newblock Orbital magnetic states in moir{\'e} graphene systems.
\newblock \emph{Nature Reviews Physics}, pages 1--16, 2021.

\bibitem[Chu et~al.(2020)Chu, Liu, Yuan, Shen, Yang, Shi, Yang, and
  Zhang]{Chu_2020}
Yanbang Chu, Le~Liu, Yalong Yuan, Cheng Shen, Rong Yang, Dongxia Shi, Wei Yang,
  and Guangyu Zhang.
\newblock A review of experimental advances in twisted graphene moir{\'{e}}
  superlattice.
\newblock \emph{Chinese Physics B}, 29\penalty0 (12):\penalty0 128104, dec
  2020.
\newblock \doi{10.1088/1674-1056/abb221}.
\newblock URL \url{https://doi.org/10.1088/1674-1056/abb221}.

\bibitem[Herzog-Arbeitman et~al.(2020)Herzog-Arbeitman, Song, Regnault, and
  Bernevig]{PhysRevLett.125.236804}
Jonah Herzog-Arbeitman, Zhi-Da Song, Nicolas Regnault, and B.~Andrei Bernevig.
\newblock Hofstadter topology: Noncrystalline topological materials at high
  flux.
\newblock \emph{Phys. Rev. Lett.}, 125:\penalty0 236804, Dec 2020.
\newblock \doi{10.1103/PhysRevLett.125.236804}.
\newblock URL \url{https://link.aps.org/doi/10.1103/PhysRevLett.125.236804}.

\bibitem[{Bistritzer} and {MacDonald}(2011)]{2011PNAS..10812233B}
Rafi {Bistritzer} and Allan~H. {MacDonald}.
\newblock {Moir{\'e} bands in twisted double-layer graphene}.
\newblock \emph{Proceedings of the National Academy of Science}, 108\penalty0
  (30):\penalty0 12233--12237, Jul 2011.
\newblock \doi{10.1073/pnas.1108174108}.

\bibitem[Koshino et~al.(2018)Koshino, Yuan, Koretsune, Ochi, Kuroki, and
  Fu]{PhysRevX.8.031087}
Mikito Koshino, Noah F.~Q. Yuan, Takashi Koretsune, Masayuki Ochi, Kazuhiko
  Kuroki, and Liang Fu.
\newblock Maximally localized wannier orbitals and the extended hubbard model
  for twisted bilayer graphene.
\newblock \emph{Phys. Rev. X}, 8:\penalty0 031087, Sep 2018.
\newblock \doi{10.1103/PhysRevX.8.031087}.
\newblock URL \url{https://link.aps.org/doi/10.1103/PhysRevX.8.031087}.

\bibitem[{Das} et~al.(2021){Das}, {Lu}, {Herzog-Arbeitman}, {Song}, {Watanabe},
  {Taniguchi}, {Bernevig}, and {Efetov}]{2020arXiv200713390D}
Ipsita {Das}, Xiaobo {Lu}, Jonah {Herzog-Arbeitman}, Zhi-Da {Song}, Kenji
  {Watanabe}, Takashi {Taniguchi}, B.~Andrei {Bernevig}, and Dmitri~K.
  {Efetov}.
\newblock {Symmetry-broken Chern insulators and Rashba-like Landau-level
  crossings in magic-angle bilayer graphene}.
\newblock \emph{Nature Physics}, 17\penalty0 (6):\penalty0 710--714, January
  2021.
\newblock \doi{10.1038/s41567-021-01186-3}.

\bibitem[{Song} et~al.(2019){Song}, {Wang}, {Shi}, {Li}, {Fang}, and
  {Bernevig}]{2018arXiv180710676S}
Zhi-Da {Song}, Zhijun {Wang}, Wujun {Shi}, Gang {Li}, Chen {Fang}, and
  B.~Andrei {Bernevig}.
\newblock {All Magic Angles in Twisted Bilayer Graphene are Topological}.
\newblock \emph{\prl}, 123\penalty0 (3):\penalty0 036401, Jul 2019.
\newblock \doi{10.1103/PhysRevLett.123.036401}.

\bibitem[Tarnopolsky et~al.(2019)Tarnopolsky, Kruchkov, and
  Vishwanath]{PhysRevLett.122.106405}
Grigory Tarnopolsky, Alex~Jura Kruchkov, and Ashvin Vishwanath.
\newblock Origin of magic angles in twisted bilayer graphene.
\newblock \emph{Phys. Rev. Lett.}, 122:\penalty0 106405, Mar 2019.
\newblock \doi{10.1103/PhysRevLett.122.106405}.
\newblock URL \url{https://link.aps.org/doi/10.1103/PhysRevLett.122.106405}.

\bibitem[{Bernevig} et~al.(2021{\natexlab{a}}){Bernevig}, {Song}, {Regnault},
  and {Lian}]{2020arXiv200911301B}
B.~Andrei {Bernevig}, Zhi-Da {Song}, Nicolas {Regnault}, and Biao {Lian}.
\newblock {Twisted bilayer graphene. I. Matrix elements, approximations,
  perturbation theory, and a k .p two-band model}.
\newblock \emph{\prb}, 103\penalty0 (20):\penalty0 205411, May
  2021{\natexlab{a}}.
\newblock \doi{10.1103/PhysRevB.103.205411}.

\bibitem[{Song} et~al.(2021){Song}, {Lian}, {Regnault}, and
  {Bernevig}]{2020arXiv200911872S}
Zhi-Da {Song}, Biao {Lian}, Nicolas {Regnault}, and B.~Andrei {Bernevig}.
\newblock {Twisted bilayer graphene. II. Stable symmetry anomaly}.
\newblock \emph{\prb}, 103\penalty0 (20):\penalty0 205412, May 2021.
\newblock \doi{10.1103/PhysRevB.103.205412}.

\bibitem[{Wang} et~al.(2020){Wang}, {Zheng}, {Millis}, and
  {Cano}]{2020arXiv201003589W}
Jie {Wang}, Yunqin {Zheng}, Andrew~J. {Millis}, and Jennifer {Cano}.
\newblock {Chiral Approximation to Twisted Bilayer Graphene: Exact Intra-Valley
  Inversion Symmetry, Nodal Structure and Implications for Higher Magic
  Angles}.
\newblock \emph{arXiv e-prints}, art. arXiv:2010.03589, October 2020.

\bibitem[{Chen} et~al.(2020){Chen}, {Da Liao}, {Chen}, {Vafek}, {Kang}, {Li},
  and {Meng}]{2020arXiv201107602C}
Bin-Bin {Chen}, Yuan {Da Liao}, Ziyu {Chen}, Oskar {Vafek}, Jian {Kang}, Wei
  {Li}, and Zi~Yang {Meng}.
\newblock {Realization of Topological Mott Insulator in a Twisted Bilayer
  Graphene Lattice Model}.
\newblock \emph{arXiv e-prints}, art. arXiv:2011.07602, November 2020.

\bibitem[Zou et~al.(2018)Zou, Po, Vishwanath, and Senthil]{PhysRevB.98.085435}
Liujun Zou, Hoi~Chun Po, Ashvin Vishwanath, and T.~Senthil.
\newblock Band structure of twisted bilayer graphene: Emergent symmetries,
  commensurate approximants, and wannier obstructions.
\newblock \emph{Phys. Rev. B}, 98:\penalty0 085435, Aug 2018.
\newblock \doi{10.1103/PhysRevB.98.085435}.
\newblock URL \url{https://link.aps.org/doi/10.1103/PhysRevB.98.085435}.

\bibitem[Hofstadter(1976)]{PhysRevB.14.2239}
Douglas~R. Hofstadter.
\newblock Energy levels and wave functions of bloch electrons in rational and
  irrational magnetic fields.
\newblock \emph{Phys. Rev. B}, 14:\penalty0 2239--2249, Sep 1976.
\newblock \doi{10.1103/PhysRevB.14.2239}.

\bibitem[Wang and Santos(2020)]{PhysRevLett.125.236805}
Jian Wang and Luiz~H. Santos.
\newblock Classification of topological phase transitions and van hove
  singularity steering mechanism in graphene superlattices.
\newblock \emph{Phys. Rev. Lett.}, 125:\penalty0 236805, Dec 2020.
\newblock \doi{10.1103/PhysRevLett.125.236805}.
\newblock URL \url{https://link.aps.org/doi/10.1103/PhysRevLett.125.236805}.

\bibitem[Albrecht et~al.(2001)Albrecht, Smet, von Klitzing, Weiss, Umansky, and
  Schweizer]{PhysRevLett.86.147}
C.~Albrecht, J.~H. Smet, K.~von Klitzing, D.~Weiss, V.~Umansky, and
  H.~Schweizer.
\newblock Evidence of hofstadter's fractal energy spectrum in the quantized
  hall conductance.
\newblock \emph{Phys. Rev. Lett.}, 86:\penalty0 147--150, Jan 2001.
\newblock \doi{10.1103/PhysRevLett.86.147}.
\newblock URL \url{https://link.aps.org/doi/10.1103/PhysRevLett.86.147}.

\bibitem[Bernevig and Hughes(2013)]{andreibook}
B.~Andrei Bernevig and Taylor~L. Hughes.
\newblock \emph{Topological Insulators and Topological Superconductors}.
\newblock Princeton University Press, student edition edition, 2013.
\newblock ISBN 9780691151755.

\bibitem[{Lu} et~al.(2020){Lu}, {Lian}, {Chaudhary}, {Piot}, {Romagnoli},
  {Watanabe}, {Taniguchi}, {Poggio}, {MacDonald}, {Bernevig}, and
  {Efetov}]{2020arXiv200613963L}
Xiaobo {Lu}, Biao {Lian}, Gaurav {Chaudhary}, Benjamin~A. {Piot}, Giulio
  {Romagnoli}, Kenji {Watanabe}, Takashi {Taniguchi}, Martino {Poggio},
  Allan~H. {MacDonald}, B.~Andrei {Bernevig}, and Dmitri~K. {Efetov}.
\newblock {Fingerprints of Fragile Topology in the Hofstadter spectrum of
  Twisted Bilayer Graphene Close to the Second Magic Angle}.
\newblock \emph{PNAS}, art. arXiv:2006.13963, June 2020.

\bibitem[Dean et~al.(2013)Dean, Wang, Maher, Forsythe, Ghahari, Gao, Katoch,
  Ishigami, Moon, Koshino, Taniguchi, Watanabe, Shepard, Hone, and
  Kim]{moirehofexp}
C.~R. Dean, L.~Wang, P.~Maher, C.~Forsythe, F.~Ghahari, Y.~Gao, J.~Katoch,
  M.~Ishigami, P.~Moon, M.~Koshino, T.~Taniguchi, K.~Watanabe, K.~L. Shepard,
  J.~Hone, and P.~Kim.
\newblock Hofstadter's butterfly and the fractal quantum hall effect in
  moir{\'e}superlattices.
\newblock \emph{Nature}, 497:\penalty0 598 EP --, 05 2013.

\bibitem[Zak(1964{\natexlab{a}})]{PhysRev.134.A1602}
J.~Zak.
\newblock Magnetic translation group.
\newblock \emph{Phys. Rev.}, 134:\penalty0 A1602--A1606, Jun
  1964{\natexlab{a}}.
\newblock \doi{10.1103/PhysRev.134.A1602}.
\newblock URL \url{https://link.aps.org/doi/10.1103/PhysRev.134.A1602}.

\bibitem[Zak(1964{\natexlab{b}})]{PhysRev.134.A1607}
J.~Zak.
\newblock Magnetic translation group. ii. irreducible representations.
\newblock \emph{Phys. Rev.}, 134:\penalty0 A1607--A1611, Jun
  1964{\natexlab{b}}.
\newblock \doi{10.1103/PhysRev.134.A1607}.
\newblock URL \url{https://link.aps.org/doi/10.1103/PhysRev.134.A1607}.

\bibitem[Brown(1969)]{BROWN1969313}
E.~Brown.
\newblock Aspects of group theory in electron dynamics**this work supported by
  the u.s. atomic energy commission.
\newblock 22:\penalty0 313--408, 1969.
\newblock ISSN 0081-1947.
\newblock \doi{https://doi.org/10.1016/S0081-1947(08)60033-8}.
\newblock URL
  \url{https://www.sciencedirect.com/science/article/pii/S0081194708600338}.

\bibitem[Streda(1982)]{Streda_1982}
P~Streda.
\newblock Theory of quantised hall conductivity in two dimensions.
\newblock \emph{Journal of Physics C: Solid State Physics}, 15\penalty0
  (22):\penalty0 L717--L721, aug 1982.
\newblock \doi{10.1088/0022-3719/15/22/005}.
\newblock URL \url{https://doi.org/10.1088/0022-3719/15/22/005}.

\bibitem[{Wannier}(1978)]{1978PSSBR..88..757W}
G.~H. {Wannier}.
\newblock {A Result Not Dependent on Rationality for Bloch Electrons in a
  Magnetic Field}.
\newblock \emph{Physica Status Solidi B Basic Research}, 88\penalty0
  (2):\penalty0 757--765, August 1978.
\newblock \doi{10.1002/pssb.2220880243}.

\bibitem[Pereira et~al.(2007)Pereira, Peeters, and
  Vasilopoulos]{PhysRevB.76.115419}
J.~Milton Pereira, F.~M. Peeters, and P.~Vasilopoulos.
\newblock Landau levels and oscillator strength in a biased bilayer of
  graphene.
\newblock \emph{Phys. Rev. B}, 76:\penalty0 115419, Sep 2007.
\newblock \doi{10.1103/PhysRevB.76.115419}.
\newblock URL \url{https://link.aps.org/doi/10.1103/PhysRevB.76.115419}.

\bibitem[Xiao et~al.(2010)Xiao, Chang, and Niu]{RevModPhys.82.1959}
Di~Xiao, Ming-Che Chang, and Qian Niu.
\newblock Berry phase effects on electronic properties.
\newblock \emph{Rev. Mod. Phys.}, 82:\penalty0 1959--2007, Jul 2010.
\newblock \doi{10.1103/RevModPhys.82.1959}.
\newblock URL \url{https://link.aps.org/doi/10.1103/RevModPhys.82.1959}.

\bibitem[Gumbs et~al.(1995)Gumbs, Miessein, and Huang]{PhysRevB.52.14755}
Godfrey Gumbs, Desir\'e Miessein, and Danhong Huang.
\newblock Effect of magnetic modulation on bloch electrons on a two-dimensional
  square lattice.
\newblock \emph{Phys. Rev. B}, 52:\penalty0 14755--14760, Nov 1995.
\newblock \doi{10.1103/PhysRevB.52.14755}.
\newblock URL \url{https://link.aps.org/doi/10.1103/PhysRevB.52.14755}.

\bibitem[Bistritzer and MacDonald(2011)]{PhysRevB.84.035440}
R.~Bistritzer and A.~H. MacDonald.
\newblock Moir\'e butterflies in twisted bilayer graphene.
\newblock \emph{Phys. Rev. B}, 84:\penalty0 035440, Jul 2011.
\newblock \doi{10.1103/PhysRevB.84.035440}.
\newblock URL \url{https://link.aps.org/doi/10.1103/PhysRevB.84.035440}.

\bibitem[{Hejazi} et~al.(2019){Hejazi}, {Liu}, and
  {Balents}]{2019PhRvB.100c5115H}
Kasra {Hejazi}, Chunxiao {Liu}, and Leon {Balents}.
\newblock {Landau levels in twisted bilayer graphene and semiclassical orbits}.
\newblock \emph{\prb}, 100\penalty0 (3):\penalty0 035115, July 2019.
\newblock \doi{10.1103/PhysRevB.100.035115}.

\bibitem[Crosse et~al.(2020)Crosse, Nakatsuji, Koshino, and Moon]{Crosse_2020}
J.~A. Crosse, Naoto Nakatsuji, Mikito Koshino, and Pilkyung Moon.
\newblock Hofstadter butterfly and the quantum hall effect in twisted double
  bilayer graphene.
\newblock \emph{Physical Review B}, 102\penalty0 (3), Jul 2020.
\newblock ISSN 2469-9969.
\newblock \doi{10.1103/physrevb.102.035421}.
\newblock URL \url{http://dx.doi.org/10.1103/PhysRevB.102.035421}.

\bibitem[{Lian} et~al.(2021{\natexlab{a}}){Lian}, {Xie}, and
  {Bernevig}]{2021PhRvB.103L1405L}
Biao {Lian}, Fang {Xie}, and B.~Andrei {Bernevig}.
\newblock {Open momentum space method for the Hofstadter butterfly and the
  quantized Lorentz susceptibility}.
\newblock \emph{\prb}, 103\penalty0 (16):\penalty0 L161405, April
  2021{\natexlab{a}}.
\newblock \doi{10.1103/PhysRevB.103.L161405}.

\bibitem[Herzog-Arbeitman et~al.()Herzog-Arbeitman, Chew, and
  Bernevig]{secondpaper}
Jonah Herzog-Arbeitman, Aaron Chew, and Andrei Bernevig.
\newblock The magnetic bloch theorem at {$2\pi$} flux.

\bibitem[Das et~al.()Das, Shen, Jaoui, Herzog-Arbeitman, Chew, Cho, Watanabe,
  Taniguchi, Piot, Bernevig, and Efetov]{exppaper}
Ipsita Das, Cheng Shen, Alexandre Jaoui, Jonah Herzog-Arbeitman, Aaron Chew,
  Chang-Woo Cho, Kenji Watanabe, Takashi Taniguchi, Benjamin~A. Piot, B.~Andrei
  Bernevig, and Dmitri~K. Efetov.
\newblock Observation of re-entrant correlated insulators and interaction
  driven fermi surface reconstructions at one magnetic flux quantum per moir\'e
  unit cell in magic-angle twisted bilayer graphene.

\bibitem[Brouder et~al.(2007)Brouder, Panati, Calandra, Mourougane, and
  Marzari]{Brouder_2007}
Christian Brouder, Gianluca Panati, Matteo Calandra, Christophe Mourougane, and
  Nicola Marzari.
\newblock Exponential localization of wannier functions in insulators.
\newblock \emph{Physical Review Letters}, 98\penalty0 (4), Jan 2007.
\newblock ISSN 1079-7114.
\newblock \doi{10.1103/physrevlett.98.046402}.

\bibitem[Alexandradinata et~al.(2014)Alexandradinata, Dai, and
  Bernevig]{Alexandradinata:2012sp}
A.~Alexandradinata, Xi~Dai, and B.~Andrei Bernevig.
\newblock {Wilson-Loop Characterization of Inversion-Symmetric Topological
  Insulators}.
\newblock \emph{Phys. Rev.}, B89\penalty0 (15):\penalty0 155114, 2014.
\newblock \doi{10.1103/PhysRevB.89.155114}.

\bibitem[{Bradlyn} et~al.(2017){Bradlyn}, {Elcoro}, {Cano}, {Vergniory},
  {Wang}, {Felser}, {Aroyo}, and {Bernevig}]{2017Natur.547..298B}
Barry {Bradlyn}, L.~{Elcoro}, Jennifer {Cano}, M.~G. {Vergniory}, Zhijun
  {Wang}, C.~{Felser}, M.~I. {Aroyo}, and B.~Andrei {Bernevig}.
\newblock {Topological quantum chemistry}.
\newblock \emph{\nat}, 547\penalty0 (7663):\penalty0 298--305, Jul 2017.
\newblock \doi{10.1038/nature23268}.

\bibitem[Aroyo et~al.(2006{\natexlab{a}})Aroyo, Perez-Mato, Capillas, Kroumova,
  Ivantchev, Madariaga, Kirov, and Wondratschek]{Aroyo:firstpaper}
MI~Aroyo, JM~Perez-Mato, Cesar Capillas, Eli Kroumova, Svetoslav Ivantchev,
  Gotzon Madariaga, Asen Kirov, and Hans Wondratschek.
\newblock Bilbao crystallographic server: I. databases and crystallographic
  computing programs.
\newblock \emph{ZEITSCHRIFT FUR KRISTALLOGRAPHIE}, 221:\penalty0 15--27, 01
  2006{\natexlab{a}}.
\newblock \doi{10.1524/zkri.2006.221.1.15}.

\bibitem[Aroyo et~al.(2006{\natexlab{b}})Aroyo, Kirov, Capillas, Perez-Mato,
  and Wondratschek]{Aroyo:xo5013}
Mois~I. Aroyo, Asen Kirov, Cesar Capillas, J.~M. Perez-Mato, and Hans
  Wondratschek.
\newblock {Bilbao Crystallographic Server. II. Representations of
  crystallographic point groups and space groups}.
\newblock \emph{Acta Crystallographica Section A}, 62\penalty0 (2):\penalty0
  115--128, Mar 2006{\natexlab{b}}.
\newblock \doi{10.1107/S0108767305040286}.

\bibitem[{Kang} and {Vafek}(2018)]{2018PhRvX...8c1088K}
Jian {Kang} and Oskar {Vafek}.
\newblock {Symmetry, Maximally Localized Wannier States, and a Low-Energy Model
  for Twisted Bilayer Graphene Narrow Bands}.
\newblock \emph{Physical Review X}, 8\penalty0 (3):\penalty0 031088, July 2018.
\newblock \doi{10.1103/PhysRevX.8.031088}.

\bibitem[{Cano} et~al.(2018{\natexlab{a}}){Cano}, {Bradlyn}, {Wang}, {Elcoro},
  {Vergniory}, {Felser}, {Aroyo}, and {Bernevig}]{2018PhRvL.120z6401C}
Jennifer {Cano}, Barry {Bradlyn}, Zhijun {Wang}, L.~{Elcoro}, M.~G.
  {Vergniory}, C.~{Felser}, M.~I. {Aroyo}, and B.~Andrei {Bernevig}.
\newblock {Topology of Disconnected Elementary Band Representations}.
\newblock \emph{\prl}, 120\penalty0 (26):\penalty0 266401, June
  2018{\natexlab{a}}.
\newblock \doi{10.1103/PhysRevLett.120.266401}.

\bibitem[{Cano} et~al.(2018{\natexlab{b}}){Cano}, {Bradlyn}, {Wang}, {Elcoro},
  {Vergniory}, {Felser}, {Aroyo}, and {Bernevig}]{2018PhRvB..97c5139C}
Jennifer {Cano}, Barry {Bradlyn}, Zhijun {Wang}, L.~{Elcoro}, M.~G.
  {Vergniory}, C.~{Felser}, M.~I. {Aroyo}, and B.~Andrei {Bernevig}.
\newblock {Building blocks of topological quantum chemistry: Elementary band
  representations}.
\newblock \emph{\prb}, 97\penalty0 (3):\penalty0 035139, Jan
  2018{\natexlab{b}}.
\newblock \doi{10.1103/PhysRevB.97.035139}.

\bibitem[Bradlyn et~al.(2019)Bradlyn, Wang, Cano, and Bernevig]{Bradlyn_2019}
Barry Bradlyn, Zhijun Wang, Jennifer Cano, and B.~Andrei Bernevig.
\newblock Disconnected elementary band representations, fragile topology, and
  wilson loops as topological indices: An example on the triangular lattice.
\newblock \emph{Physical Review B}, 99\penalty0 (4), Jan 2019.
\newblock ISSN 2469-9969.
\newblock \doi{10.1103/physrevb.99.045140}.
\newblock URL \url{http://dx.doi.org/10.1103/PhysRevB.99.045140}.

\bibitem[{Fang} et~al.(2012){Fang}, {Gilbert}, and
  {Bernevig}]{2012PhRvB..86k5112F}
Chen {Fang}, Matthew~J. {Gilbert}, and B.~Andrei {Bernevig}.
\newblock {Bulk topological invariants in noninteracting point group symmetric
  insulators}.
\newblock \emph{\prb}, 86\penalty0 (11):\penalty0 115112, September 2012.
\newblock \doi{10.1103/PhysRevB.86.115112}.

\bibitem[Liu et~al.(2019)Liu, Liu, and Dai]{PhysRevB.99.155415}
Jianpeng Liu, Junwei Liu, and Xi~Dai.
\newblock Pseudo landau level representation of twisted bilayer graphene: Band
  topology and implications on the correlated insulating phase.
\newblock \emph{Phys. Rev. B}, 99:\penalty0 155415, Apr 2019.
\newblock \doi{10.1103/PhysRevB.99.155415}.
\newblock URL \url{https://link.aps.org/doi/10.1103/PhysRevB.99.155415}.

\bibitem[{Ahn} et~al.(2019){Ahn}, {Park}, and {Yang}]{2019PhRvX...9b1013A}
J.~{Ahn}, S.~{Park}, and B.-J. {Yang}.
\newblock {Failure of Nielsen-Ninomiya Theorem and Fragile Topology in
  Two-Dimensional Systems with Space-Time Inversion Symmetry: Application to
  Twisted Bilayer Graphene at Magic Angle}.
\newblock \emph{Physical Review X}, 9\penalty0 (2):\penalty0 021013, April
  2019.
\newblock \doi{10.1103/PhysRevX.9.021013}.

\bibitem[{Bernevig} et~al.(2021{\natexlab{b}}){Bernevig}, {Song}, {Regnault},
  and {Lian}]{2020arXiv200912376B}
B.~Andrei {Bernevig}, Zhi-Da {Song}, Nicolas {Regnault}, and Biao {Lian}.
\newblock {Twisted bilayer graphene. III. Interacting Hamiltonian and exact
  symmetries}.
\newblock \emph{\prb}, 103\penalty0 (20):\penalty0 205413, May
  2021{\natexlab{b}}.
\newblock \doi{10.1103/PhysRevB.103.205413}.

\bibitem[{Liu} et~al.(2021){Liu}, {Wang}, {Watanabe}, {Taniguchi}, {Vafek}, and
  {Li}]{2021Sci...371.1261L}
Xiaoxue {Liu}, Zhi {Wang}, K.~{Watanabe}, T.~{Taniguchi}, Oskar {Vafek}, and
  J.~I.~A. {Li}.
\newblock {Tuning electron correlation in magic-angle twisted bilayer graphene
  using Coulomb screening}.
\newblock \emph{Science}, 371\penalty0 (6535):\penalty0 1261--1265, March 2021.
\newblock \doi{10.1126/science.abb8754}.

\bibitem[{Vafek} and {Kang}(2020)]{2020arXiv200909413V}
Oskar {Vafek} and Jian {Kang}.
\newblock {Towards the hidden symmetry in Coulomb interacting twisted bilayer
  graphene: renormalization group approach}.
\newblock \emph{arXiv e-prints}, art. arXiv:2009.09413, September 2020.

\bibitem[Wilson et~al.(2020)Wilson, Fu, Das~Sarma, and
  Pixley]{PhysRevResearch.2.023325}
Justin~H. Wilson, Yixing Fu, S.~Das~Sarma, and J.~H. Pixley.
\newblock Disorder in twisted bilayer graphene.
\newblock \emph{Phys. Rev. Research}, 2:\penalty0 023325, Jun 2020.
\newblock \doi{10.1103/PhysRevResearch.2.023325}.
\newblock URL \url{https://link.aps.org/doi/10.1103/PhysRevResearch.2.023325}.

\bibitem[{Parker} et~al.(2020){Parker}, {Soejima}, {Hauschild}, {Zaletel}, and
  {Bultinck}]{2020arXiv201209885P}
Daniel~E. {Parker}, Tomohiro {Soejima}, Johannes {Hauschild}, Michael~P.
  {Zaletel}, and Nick {Bultinck}.
\newblock {Strain-induced quantum phase transitions in magic angle graphene}.
\newblock \emph{arXiv e-prints}, art. arXiv:2012.09885, December 2020.

\bibitem[{Padhi} et~al.(2020){Padhi}, {Tiwari}, {Neupert}, and
  {Ryu}]{2020arXiv200502406P}
Bikash {Padhi}, Apoorv {Tiwari}, Titus {Neupert}, and Shinsei {Ryu}.
\newblock {Transport across twist angle domains in moir{\'e} graphene}.
\newblock \emph{arXiv e-prints}, art. arXiv:2005.02406, May 2020.

\bibitem[{Uri} et~al.(2020){Uri}, {Grover}, {Cao}, {Crosse}, {Bagani},
  {Rodan-Legrain}, {Myasoedov}, {Watanabe}, {Taniguchi}, {Moon}, {Koshino},
  {Jarillo-Herrero}, and {Zeldov}]{2020Natur.581...47U}
A.~{Uri}, S.~{Grover}, Y.~{Cao}, J.~{\^A}.~A. {Crosse}, K.~{Bagani},
  D.~{Rodan-Legrain}, Y.~{Myasoedov}, K.~{Watanabe}, T.~{Taniguchi}, P.~{Moon},
  M.~{Koshino}, P.~{Jarillo-Herrero}, and E.~{Zeldov}.
\newblock {Mapping the twist-angle disorder and Landau levels in magic-angle
  graphene}.
\newblock \emph{\nat}, 581\penalty0 (7806):\penalty0 47--52, May 2020.
\newblock \doi{10.1038/s41586-020-2255-3}.

\bibitem[{Kazmierczak} et~al.(2020){Kazmierczak}, {Van Winkle}, {Ophus},
  {Bustillo}, {Brown}, {Carr}, {Ciston}, {Taniguchi}, {Watanabe}, and {Kwabena
  Bediako}]{2020arXiv200809761K}
Nathanael~P. {Kazmierczak}, Madeline {Van Winkle}, Colin {Ophus}, Karen~C.
  {Bustillo}, Hamish~G. {Brown}, Stephen {Carr}, Jim {Ciston}, Takashi
  {Taniguchi}, Kenji {Watanabe}, and D.~{Kwabena Bediako}.
\newblock {Strain fields in twisted bilayer graphene}.
\newblock \emph{arXiv e-prints}, art. arXiv:2008.09761, August 2020.

\bibitem[Benschop et~al.(2021)Benschop, de~Jong, Stepanov, Lu, Stalman, van~der
  Molen, Efetov, and Allan]{PhysRevResearch.3.013153}
Tjerk Benschop, Tobias~A. de~Jong, Petr Stepanov, Xiaobo Lu, Vincent Stalman,
  Sense~Jan van~der Molen, Dmitri~K. Efetov, and Milan~P. Allan.
\newblock Measuring local moir\'e lattice heterogeneity of twisted bilayer
  graphene.
\newblock \emph{Phys. Rev. Research}, 3:\penalty0 013153, Feb 2021.
\newblock \doi{10.1103/PhysRevResearch.3.013153}.
\newblock URL \url{https://link.aps.org/doi/10.1103/PhysRevResearch.3.013153}.

\bibitem[Kang and Vafek(2019)]{PhysRevLett.122.246401}
Jian Kang and Oskar Vafek.
\newblock Strong coupling phases of partially filled twisted bilayer graphene
  narrow bands.
\newblock \emph{Phys. Rev. Lett.}, 122:\penalty0 246401, Jun 2019.
\newblock \doi{10.1103/PhysRevLett.122.246401}.
\newblock URL \url{https://link.aps.org/doi/10.1103/PhysRevLett.122.246401}.

\bibitem[{Bernevig} et~al.(2021{\natexlab{c}}){Bernevig}, {Lian}, {Cowsik},
  {Xie}, {Regnault}, and {Song}]{2020arXiv200914200B}
B.~Andrei {Bernevig}, Biao {Lian}, Aditya {Cowsik}, Fang {Xie}, Nicolas
  {Regnault}, and Zhi-Da {Song}.
\newblock {Twisted bilayer graphene. V. Exact analytic many-body excitations in
  Coulomb Hamiltonians: Charge gap, Goldstone modes, and absence of Cooper
  pairing}.
\newblock \emph{\prb}, 103\penalty0 (20):\penalty0 205415, May
  2021{\natexlab{c}}.
\newblock \doi{10.1103/PhysRevB.103.205415}.

\bibitem[{Sheffer} and {Stern}(2021)]{2021arXiv210610650S}
Yarden {Sheffer} and Ady {Stern}.
\newblock {Chiral Magic-Angle Twisted Bilayer Graphene in a Magnetic Field:
  Landau Level Correspondence, Exact Wavefunctions and Fractional Chern
  Insulators}.
\newblock \emph{arXiv e-prints}, art. arXiv:2106.10650, June 2021.

\bibitem[{Lian} et~al.(2021{\natexlab{b}}){Lian}, {Song}, {Regnault}, {Efetov},
  {Yazdani}, and {Bernevig}]{2020arXiv200913530L}
Biao {Lian}, Zhi-Da {Song}, Nicolas {Regnault}, Dmitri~K. {Efetov}, Ali
  {Yazdani}, and B.~Andrei {Bernevig}.
\newblock {Twisted bilayer graphene. IV. Exact insulator ground states and
  phase diagram}.
\newblock \emph{\prb}, 103\penalty0 (20):\penalty0 205414, May
  2021{\natexlab{b}}.
\newblock \doi{10.1103/PhysRevB.103.205414}.

\bibitem[{Lian} et~al.(2018){Lian}, {Xie}, and {Bernevig}]{2018arXiv181111786L}
Biao {Lian}, Fang {Xie}, and B.~Andrei {Bernevig}.
\newblock {The Landau Level of Fragile Topology}.
\newblock \emph{arXiv e-prints}, art. arXiv:1811.11786, November 2018.

\bibitem[{Kang} et~al.(2021){Kang}, {Bernevig}, and
  {Vafek}]{2021arXiv210401145K}
Jian {Kang}, B.~Andrei {Bernevig}, and Oskar {Vafek}.
\newblock {Cascades between light and heavy fermions in the normal state of
  magic angle twisted bilayer graphene}.
\newblock \emph{arXiv e-prints}, art. arXiv:2104.01145, April 2021.

\bibitem[{Chaudhary} et~al.(2021){Chaudhary}, {MacDonald}, and
  {Norman}]{2021arXiv210501243C}
Gaurav {Chaudhary}, A.~H. {MacDonald}, and M.~R. {Norman}.
\newblock {Quantum Hall Superconductivity from Moir\{{\'e}\} Landau Levels}.
\newblock \emph{arXiv e-prints}, art. arXiv:2105.01243, May 2021.

\bibitem[{Cao} et~al.(2021){Cao}, {Park}, {Watanabe}, {Taniguchi}, and
  {Jarillo-Herrero}]{2021arXiv210312083C}
Yuan {Cao}, Jeong~Min {Park}, Kenji {Watanabe}, Takashi {Taniguchi}, and Pablo
  {Jarillo-Herrero}.
\newblock {Large Pauli Limit Violation and Reentrant Superconductivity in
  Magic-Angle Twisted Trilayer Graphene}.
\newblock \emph{arXiv e-prints}, art. arXiv:2103.12083, March 2021.

\end{thebibliography}
\bibliographystyle{aipnum4-1}
\bibliographystyle{unsrtnat}
\let\addcontentsline\oldaddcontentsline

\setcitestyle{numbers,square}

\onecolumngrid
\appendix

\section{Magnetic Bloch Theorem Formulae}
\label{app:formula}

This Appendix includes formulae for the band structure, Wilson loop, and many-body form factors. The derivation of these results is direct but technical, and they are left to a separate work \cite{secondpaper}.

The starting point of all results in this section are the basis states
\bea
\label{eq:MTGirreps}
\ket{\mbf{k}, n, \al, l} = \frac{1}{\sqrt{\mathcal{N}(\mbf{k})}} \sum_{\mbf{R}} e^{-i \mbf{k} \cdot \mbf{R}} T_{\mbf{a}_1}^{\mbf{R} \cdot \mbf{b}_1} T_{\mbf{a}_2}^{\mbf{R} \cdot \mbf{b}_2} \ket{n, \al, l}, \qquad \mathcal{N}(\mbf{k}) = \sqrt{2} \left| \th_3\lp \left. \frac{k}{2\pi} \right| i \rp \th_3 \left. \lp \frac{i \bar{k}}{2\pi} \right|  i \rp \right| \exp \lp - \frac{k \bar{k}}{4\pi} \rp
\eea
which are magnetic translation group eigenstates (in any gauge). Here $k = k_1 + i k_2, \bar{k} = k_1 - i k_2$ and $\th_3(z|\tau)=\th_1(z+1|\tau)$ is the Jacobi theta function with quasi-period $\tau$ and zeros at $1/2 + \tau/2$. The states in \Eq{eq:MTGirreps} carry a $T_{\mbf{a}_i}$ ``momentum" quantum number, and have indices $n,\al,l$ corresponding to Landau level, sublattice, and layer. By computing the matrix elements in Eq. 7 of the Main Text, we arrive at an expression for the magnetic Bloch Hamiltonian at $2\pi$ flux:
\bea
\label{eq:magblochham}
H^{\phi = 2\pi}(\mbf{k}) = \bpm  v_F k_\th (\sqrt{\frac{\phi}{2\pi}}  h(\pmb{\pi}) - \frac{1}{2}\sigma_2) &  T_1 + T_2 e^{-i k_2} \mathcal{H}^{2\pi\mbf{b}_1} + T_3  e^{i k_1} \mathcal{H}^{2\pi\mbf{b}_2} \\
T_1 + T_2 e^{i k_2} \mathcal{H}^{-2\pi\mbf{b}_1} + T_3  e^{-i k_1} \mathcal{H}^{-2\pi\mbf{b}_2}  & v_F k_\th (\sqrt{\frac{\phi}{2\pi}}  h(\pmb{\pi})  + \frac{1}{2} \sigma_2)
\epm, \quad h(\pmb{\pi}) =   \lp\frac{3\sqrt{3}}{2\pi} \rp^{1/2}  \bpm
0 & a^\dag \\
a& 0 \\
\epm
\eea
where $T_i$ and $\sigma_i$ act on the sublattice indices (an expression for $T_i$ is given in the Main Text) and $\mathcal{H}^{2\pi \mbf{G}}$ and $a, a^\dag$ act on the Landau level basis. Explicitly:
\bea
\null [a]_{mn} &= \sqrt{n} \delta_{m,n-1}\\
\mathcal{H}^{\mbf{q}}_{mn} &=  [\exp \lp i \eps_{ij} q_i \tilde{Z}_j \rp]_{mn},
\qquad  [Z_j]_{mn}= \frac{\bar{z}_j \sqrt{n} \delta_{m,n-1} + z_j \sqrt{m} \delta_{n,m-1}}{\sqrt{2 \phi}}
\eea
with $\phi = 2\pi$ and $\bar{z}_i = (\hat{x} - i \hat{y}) \cdot \mbf{a}_i /\sqrt{|\mbf{a}_1 \times \mbf{a}_2|}$. \Eq{eq:magblochham} is the Hamiltonian in the \emph{graphene} K valley. The Hamiltonian in the graphene $K'$ valley is related by $C_{2z} H_{K}(\mbf{k}) C_{2z}^\dag = H_{K'}(-\mbf{k}) $  where $C_{2z} = \tau_0 \sigma_1 (-1)^{a^\dag a}$ and has the same spectrum. Here $\tau_0$ denotes the layer indices which are in matrix notation in \Eq{eq:magblochham}.

We now analyze the many-body Hamiltonian with the Coulomb interaction 
\bea
\label{eq:paramapp}
V(\mbf{q}) = \pi \xi^2 U_\xi \frac{\tanh \xi |\mbf{q}|/ 2}{\xi |\mbf{q}|/ 2}, \qquad  U_{\xi} = \frac{e^2}{\eps \xi} \text{ (in Gauss units)}
\eea
where $\xi \in (10,20)$nm is the screening length given by the distance between the sample gates and $\eps \sim 6$ is the dielectric of hexagonal boron nitride. At $\xi = 15$, $U_{\xi} = 17.3$meV. We need to compute the form factor ($M,N = \pm1$ index the flat bands)
\bea
\label{eq:formfactorMTBG}
M^\eta_{MN}(\mbf{k},\mbf{q}) \equiv e^{i \xi_\mbf{q}(\mbf{k})} [U_\eta^\dag(\mbf{k}-\mbf{q})  \mathcal{H}^{\mbf{q}} U_\eta(\mbf{k})]_{MN} 
\eea
where $U_\eta(\mbf{k})$ is the matrix of occupied eigenvectors in the $\eta = K,K'$ graphene valleys and 
\bea
e^{i \xi_\mbf{q}(\mbf{k})} &= \frac{e^{- \frac{ \bar{q} q}{4\phi}}  \vartheta \lp \left. \frac{(k_1 - q/2 ,k_2+ i q/2)}{2\pi} \right| \Phi \rp  }{\sqrt{\vartheta \lp \left. \frac{(k_1 ,k_2)}{2\pi} \right| \Phi \rp \vartheta \lp \left. \frac{(k_1 - q_1 ,k_2 - q_2)}{2\pi} \right| \Phi \rp}}, \quad \Phi =  \frac{i \phi}{4\pi} \bpm 1 & i \\  i & 1 \epm
\eea
and the Siegel (or Riemann) theta function is defined
\bea
\vartheta \lp \mbf{z} \left| A \right. \rp = \sum_{\mbf{n}\in\mathds{Z}^2} e^{2\pi i \lp \frac{1}{2} \mbf{n} \cdot A \cdot \mbf{n}- \mbf{z} \cdot \mbf{n} \rp} \ . \\
\eea

\section{Additional Band Structure Plots}
\label{app:moreplots}

This Appendix includes additional plots which support some peripheral claims in the Main Text.

\begin{figure*}[h]
\centering
\includegraphics[height = 5cm]{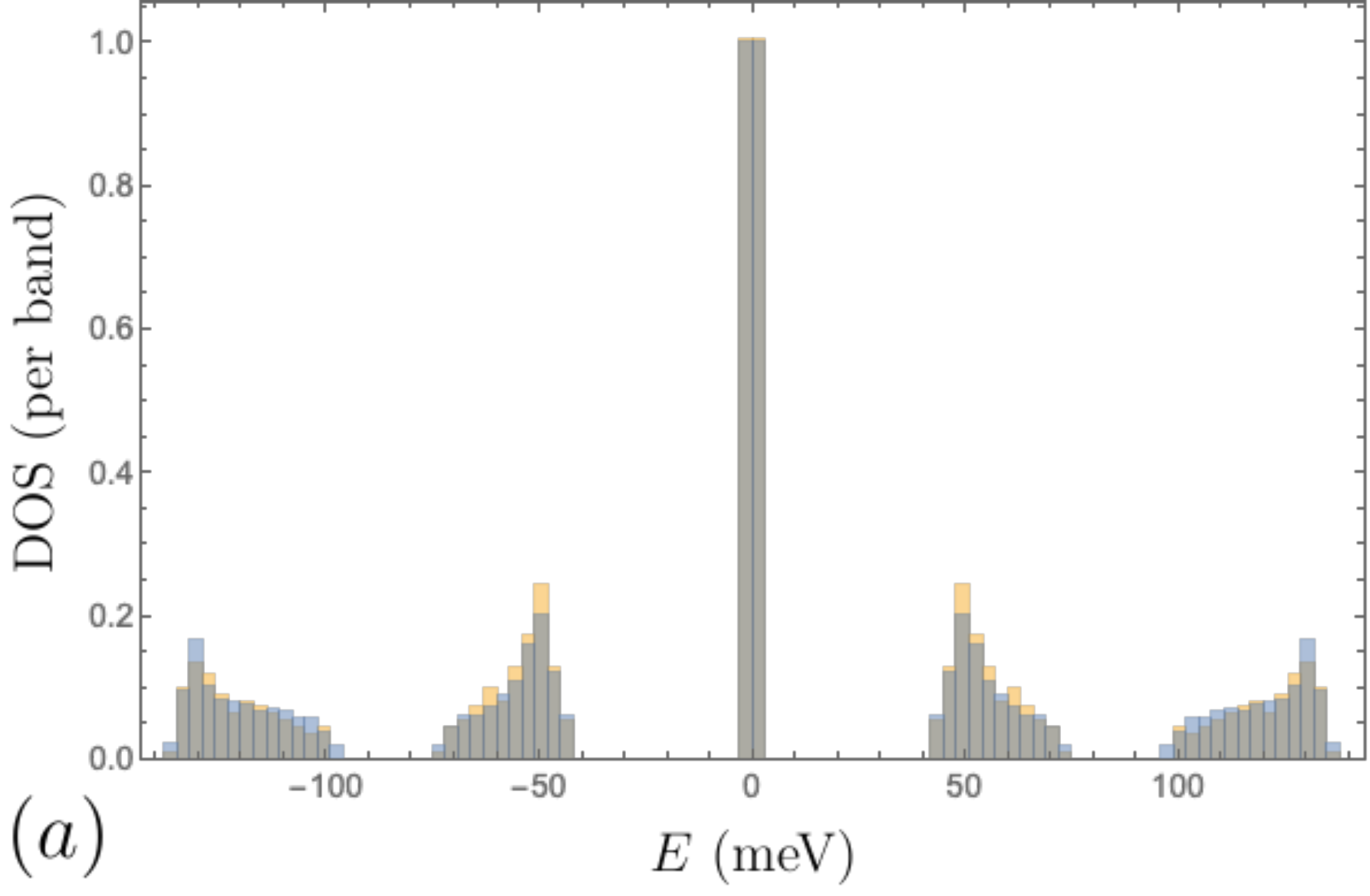} \quad
\includegraphics[height = 5cm]{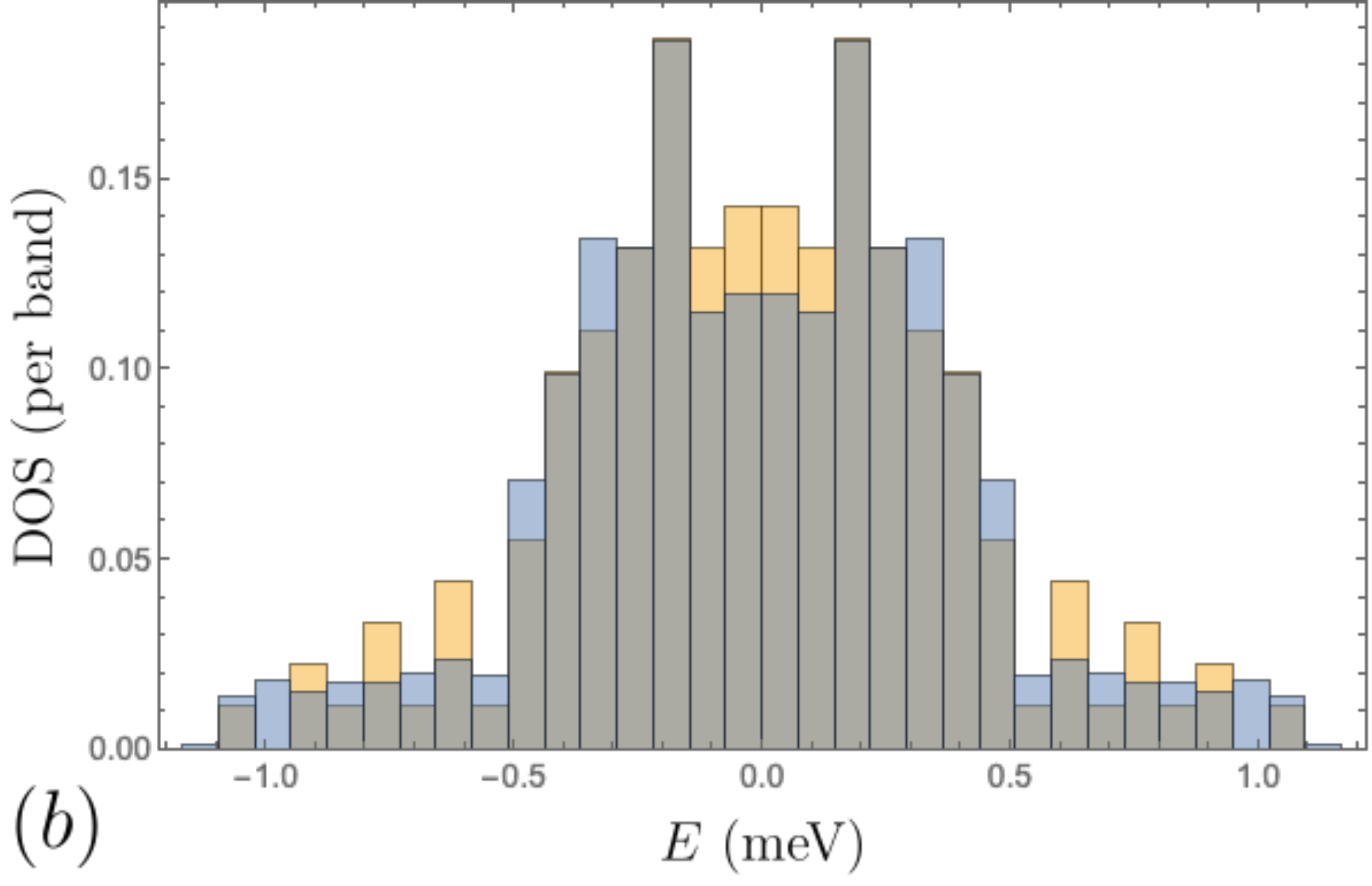}
\caption{Density of states comparison calculated using the open momentum space method (blue) and exact band structure method (yellow). The overlaps are colored gray. $(a)$  Comparison over the lowest six bands. $(b)$ Zoom in of the flat bands between $\pm1$meV. In both cases, the agreement is very good. Further improvement can be achieved by increasing the number of momentum space plaquettes in the open momentum space method. In the $84600\times 84600$ matrix, each band has approximately $122$ states \cite{2021PhRvB.103L1405L} (before spurious states are projected out) compared to the band structure technique which we use to compute $1000$ states per band. To achieve convergence between the two densities, more points per band are required.}
\label{fig:doscompar}
\end{figure*}

In \Fig{fig:doscompar}, we compare the density of states calculated using two methods, the open momentum space sparse matrix approach developed in \Ref{2021PhRvB.103L1405L} and the exact band structure approach. We can only compare the density of states between the two methods because the open momentum space approach does not keep the $\mbf{k}$ quantum number. (The advantage of the momentum space approach is a sparse  matrix representation at \emph{all} values of $\phi$.) To find quantitative agreement over a $\sim 1$meV scale, we need to use a very large sparse matrix, keeping $141$ momentum space sites in each layer and $150$ Landau levels for an $84600\times 84600$ matrix. We calculate the lowest thousand eigenvalues with the Arnoldi algorithm and employ the projector technique described in \Ref{2021PhRvB.103L1405L} at 122 momentum space plaquettes to remove the spurious states. For comparison, we only need to keep $50$ Landau levels per sublattice per layer in the band structure method, and we sample $\sim 1000$ $\mbf{k}$ points in the BZ for high accuracy. This calculation takes less than a minute.

\begin{figure*}[h]
\centering
\includegraphics[height = 6cm]{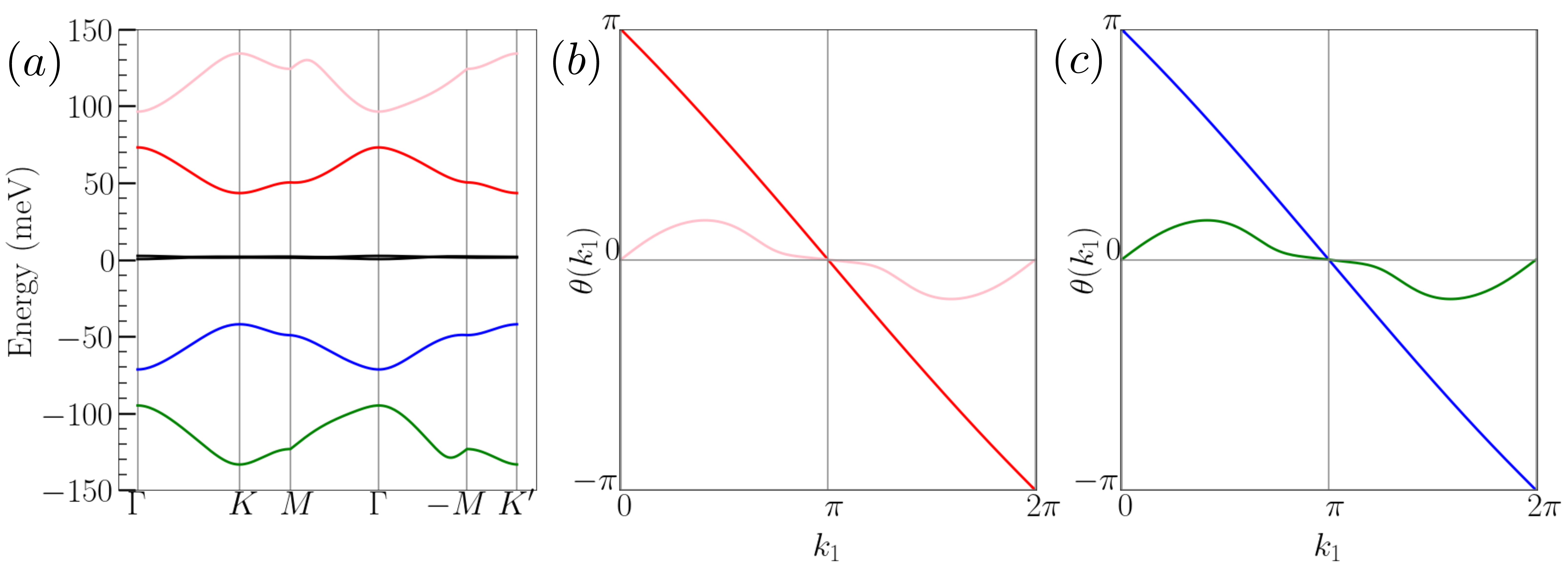}
\caption{Wilson loops of the passive bands. In $(a)$, we show the $2\pi$ flux band structure at $\th = 1.05^\circ$ with the inclusion of particle-hole breaking terms (see \App{app:PH}). In $(b), (c)$, we compute the Wilson loop over individual bands which are color coded to match $(a)$. We determine from the winding that the first passive passive bands (red and blue) have Chern number $-1$, and the second passive bands (which are gapped) have Chern number 0.}
\label{fig:passWL}
\end{figure*}

In \Fig{fig:passWL}, we study the topology of the low lying passive bands. The first important observation is that the passive bands are gapped from each other and the flat bands. This is not the case in zero flux TBG where the first and second passive bands are connected \cite{2018arXiv180710676S} with a Rashba-like dispersion at the $\Gamma$ point \cite{2020arXiv200713390D} due to $C_{3z}$ and $C_{2z}\mathcal{T}$. Recalling Fig. 1b of the Main Text, we see that the first passive bands at $2\pi$ flux (colored red and blue in \Fig{fig:passWL}a) originate from a Landau level which grows linearly in $B$ at small flux. This is exactly what is predicted from the Rashba point discussed in \Ref{2020arXiv200713390D}. As the flux increases, the Landau level degeneracy is broken due to dispersion, acquiring a bandwidth of $\sim 30$meV at $2\pi$ flux. We calculate the Chern number of the bands in \Fig{fig:passWL}b,c and confirm that the first passive bands have $C=-1$, which is the Chern number of a Landau level in our conventions (see \Ref{secondpaper} for a direct calculation). We wish to point out an essential difference between the Chern number topology of the flat bands and the Chern numbers of the passive bands. The latter are simply Landau levels (magnetic field induced topology) that have split from the passive bands at $\phi \neq 0$, while the former are a split elementary band representation (crystalline topology) which cannot be described by decoupled Landau levels. Lastly, we calculate the Chern numbers of the second passive bands, and find that they are trivial, i.e. they represent atomic states despite the strong flux.
The active bands have Chern number $C = \pm 1$, as the elementary band representation splitting in Eq. 11 of the Main Text shows. Normally, a band with Chern number $C = +1$ can annihilate the topology of a band with $C = -1$ by switching the Chern number in a phase transition.  However, in our case, these phase transitions cannot happen:  at the high-symmetry momentum $\Gamma$ we have avoided crossings.  If the transitions happen at generic points in the band structure, they will happen in pairs (because of $MT$ symmetry), and those pairs come in triplets (because of $C_{3z}$ symmetry), so the Chern number will change by $6$ and this cannot turn the Chern number of the bands to $0$.

\section{Particle-Hole Breaking Terms}
\label{app:PH}

In this Appendix, we provide details for incorporating the small angle corrections to the kinetic term of the BM model (Eq. 1 of the Main Text) into the magnetic Bloch Hamiltonian and numerically calculate the Wilson loop.  As shown in Sec. III of the Main Text, the band representation $\mathcal{B}$ of the flat bands is a decomposable elementary band representation induced from atomic orbitals. While the two flat bands are connected (as is enforced by $PM\mathcal{T}$), the topology is trivial, as we calculated directly with the Wilson loop. We now show that $O(\th)$ terms arising from the relative twist in the kinetic term of BM model \cite{2018arXiv180710676S} break the anti-commuting $P$ symmetry, gapping the flat bands are decomposing $\mathcal{B}$ into disconnected bands of opposite Chern number.

To verify this topology numerically, we study the BM Hamiltonian \emph{without} the particle-hole symmetric approximation. As written in \Ref{2020arXiv200911872S}, the Hamiltonian takes the form
\begin{align}
H_{BM,\th} = \begin{pmatrix}
-iv_F (\pmb{\nabla} \cdot \pmb{\sigma} - \frac{\theta}{2} \pmb{\nabla} \times \pmb{\sigma}) & T^\dag(\mbf{r}) \\
T(\mbf{r}) & -iv_F (\pmb{\nabla} \cdot \pmb{\sigma} + \frac{\theta}{2} \pmb{\nabla} \times \pmb{\sigma}).
\end{pmatrix}
\label{eq:HBMth}
\end{align}
which is identical to the expression in Eq. 1 of the Main Text with the addition of the $\pm \frac{\th}{2} \pmb{\nabla} \times \pmb{\sigma}$ terms which incorporate the opposite rotation of the kinetic terms in the top and bottom layers. Letting $\tau_i$ denote Pauli matrices acting on the layer index (which is the matrix notation in \Eq{eq:HBMth}), the additional term is $H_{\th} \equiv i v_F \frac{\th}{2} \tau_3 \pmb{\nabla} \times \pmb{\sigma}$. It is direct to see that $H_{\th}$ breaks particle-hole symmetry $P$ which obeys $P H_{BM}(\mbf{r}) P^\dag = - H_{BM}(\mbf{r})$. Using the expression for $P = i \tau_2 R_\pi$ where $R_\pi$ is the $\pi$ rotation operator on functions (see \Ref{secondpaper}), we find that $P H_{\th}(\mbf{r}) P^\dag = + H_{\th}(\mbf{r})$, breaking particle-hole symmetry.  We remark that at zero flux, the topology and spectrum of the BM model is not strongly influenced by $P$ because $C_{2z}\mathcal{T}$ ensures the connectedness of the flat bands and protects their topology. The $O(\th)$ terms which break $P$ at $2\pi$ flux have a more significant effect because $C_{2z}\mathcal{T}$ is also broken, allowing the $O(\th)$ particle-hole breaking terms to open a gap between the flat bands.

\begin{figure*}[h]
\centering
\includegraphics[height = 4.5cm]{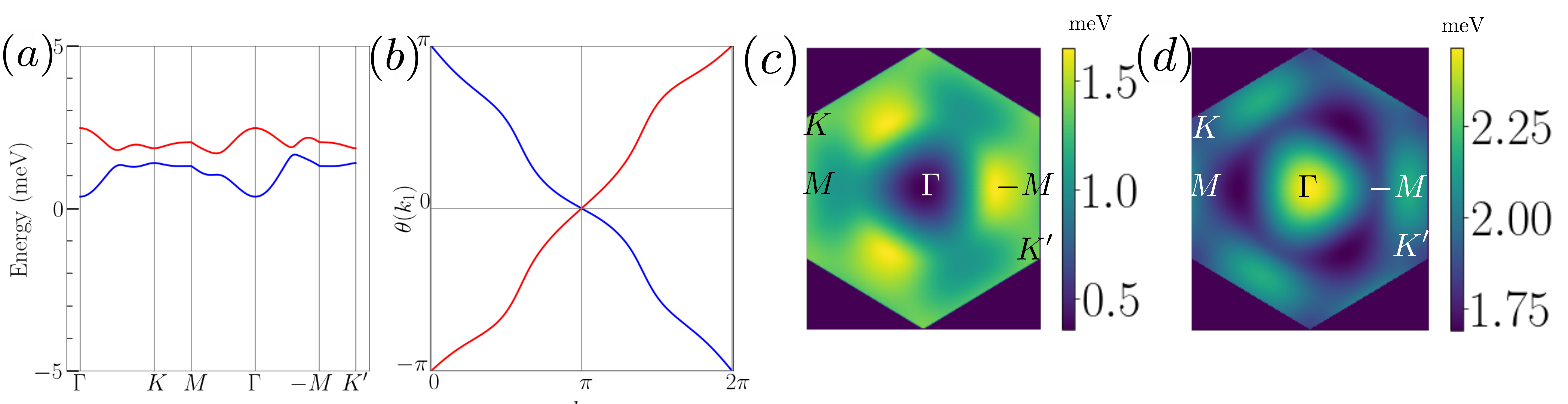}
\caption{TBG flat bands at $2\pi$ flux with particle-hole breaking terms. $(a)$ Examining the band structure reveals gaps opened at the $K$ and $K'$ points, allowing the flat bands to separate. This is only possible in flux when $C_{2z}\mathcal{T}$ is broken. $(b)$ The Abelian Wilson loop, calculated over the two bands separately, reproduces the $C= \pm1$ Chern numbers predicted from the split elementary band representation. $(c)$ and $(d)$ show the band structure across the BZ of the lower and upper flat bands respectively. The $C_{3z}$ and $M\mathcal{T}$ symmetries are evident in the spectrum. }
\label{fig:PHbreakingBS}
\end{figure*}

We now discuss the form of $H_\th$ at $2\pi$ flux. We perform the canonical substitution $-i \pmb{\nabla} \rightarrow {\pmb{\pi}}$ to find
\begin{align}
H_{BM,\th}^\phi(\mbf{r}) &= \begin{pmatrix}
v_F (\pmb{\pi} \cdot \pmb{\sigma} - \frac{\theta}{2} \pmb{\pi} \times \pmb{\sigma}) & T^\dag(\mbf{r}) \\
T(\mbf{r}) & v_F (\pmb{\pi} \cdot \pmb{\sigma} + \frac{\theta}{2} \pmb{\pi} \times \pmb{\sigma}).
\end{pmatrix}
\label{eq:HBMth}
\end{align}
As written, $H_{BM,\th}^\phi(\mbf{r})$ is not in Bloch form because $H_{BM,\th}^\phi(\mbf{r}+\mbf{a}_i) \neq H_{BM,\th}^\phi(\mbf{r})$. To remedy this, we shift into Bloch form via the unitary transformation:
\begin{align}
V_1 = \begin{pmatrix}
e^{i\pi \mbf{q}_1 \cdot \mbf{r}} & 0 \\ 0 & e^{-i\pi \mbf{q}_1 \cdot \mbf{r}}
\end{pmatrix}
\label{}
\end{align}
which acts as a momentum shift in each layer, reflecting the fact that the Dirac points in the two layers are displaced $2\pi \mbf{q}_1$ from each other.  In this section, we only discuss the particle-hole breaking term $H_\th$. All other terms are given explicitly in \App{app:formula}. We compute
\begin{align}
V_1 H_\th^\phi V_1^\dag &= -v_F \begin{pmatrix}
\frac{\theta}{2} \pmb{\pi} \times \pmb{\sigma} & 0 \\
0 & -\frac{\theta}{2} \pmb{\pi} \times \pmb{\sigma}.
\end{pmatrix} + \pi v_F \begin{pmatrix}
\frac{\theta}{2}   \mbf{q}_1 \times \pmb{\sigma} & 0 \\
0 & \frac{\theta}{2} \mbf{q}_1 \times \pmb{\sigma}
\end{pmatrix} \\
&= -v_F\frac{\theta}{2} \tau_3 \pmb{\pi} \times \pmb{\sigma} + \pi v_F  \frac{\theta}{2}  \tau_0
  \mbf{q}_1 \times \pmb{\sigma} \ .
\end{align}
The new term arising from the twist in the kinetic energy is $-v_F\frac{\theta}{2} \tau_3 \pmb{\pi} \times \pmb{\sigma} = -v_F\frac{\theta}{2} \tau_3 (\pi_x \sigma_y - \pi_y \sigma_x)$. Expanding the $\sigma$ matrices, we get
\bea
\pi_x \sigma_y - \pi_y \sigma_x &= \begin{pmatrix}
0 & -i(\pmb{\pi}_x - i \pmb{\pi}_y) \\
i(\pmb{\pi}_x + i \pmb{\pi}_y) & 0
\end{pmatrix} = \sqrt{2B} \begin{pmatrix}
0 & -i a^\dagger \\
i a & 0
\end{pmatrix}
\eea
which acts on the sublattice vector indices and the Landau level indices. Making use of $B = \phi/\Omega$ and $\Omega = \frac{2}{3\sqrt{3}} (2\pi/k_\th)^2$ from Sec. I of the Main Text, we arrive at
\bea
\label{eq:Hthfinal}
V_1 H_\th^\phi V_1^\dag &= \frac{\theta}{2} v_F k_\theta \sqrt{\dfrac{3\sqrt{3}}{2\pi}} \tau_z \otimes \begin{pmatrix}
0 & i a^\dagger \\
-i a & 0
\end{pmatrix} -\frac{\th}{4} v_F k_\th \tau_0 \sigma_x
\eea
where we also use $2\pi\mbf{q}_1 \times \pmb{\sigma}  = - k_\theta \sigma_x$. The second term in \Eq{eq:Hthfinal} acts trivially on the Landau level indices. Both terms in \Eq{eq:Hthfinal} are smaller than the leading order kinetic term by a factor of $\th$. Calculating the matrix elements of \Eq{eq:Hthfinal} on the magnetic translation operator basis states $\ket{\mbf{k},n, \al, l}$ (Eq. 6 of the Main Text), we compute the band structure. As shown in \Fig{fig:PHbreakingBS}a, the Dirac points at $K$ and $K'$ open, leaving the flat bands gapped from each other. We calculate the Wilson loop over each band individually and confirm the $C = \pm 1$ Chern numbers of the split elementary band representation in \Fig{fig:PHbreakingBS}b. Lastly, we calculate the dispersion of the flat bands over the full BZ in \Fig{fig:PHbreakingBS}c,d.

\section{Charge $\pm1$ Excitation Spectrum}
\label{app:BMchargeR}

In this section, we give a self-contained derivation of the effective Hamiltonians for charge $\pm1$ excitations following the method of \Ref{2020arXiv200913530L}. The exact eigenstates $\ket{\Psi_\nu}$ defined in the Main Text are amenable to the calculation of various excitation spectra \cite{2020arXiv200913530L}. We describe the simplest case of charge $\pm1$ excitations, which correspond to adding or removing a single electron from $\ket{\Psi_\nu}$. As discussed at length in \Ref{secondpaper}, the interaction Hamiltonian is
\bea
H_{int} = \frac{1}{2 \Omega_{tot}} \sum_{\mbf{q}} V(\mbf{q}) \bar\rho_{-\mbf{q}} \bar\rho_{\mbf{q}} &=  \frac{1}{2  \Omega_{tot}} \sum_{\mbf{G}} \sum_{\mbf{q} \in BZ} O_{-\mbf{q},-\mbf{G}} O_{\mbf{q}, \mbf{G}}, \\
 O_{\mbf{q}, \mbf{G}} &= \sqrt{V(\mbf{q}+2\pi \mbf{G})} \sum_{\mbf{k}\in BZ} \sum_{\eta, s} \sum_{MN} \bar{M}^\eta_{MN}(\mbf{k},\mbf{q}+2\pi \mbf{G}) (\gamma^\dag_{\mbf{k}- \mbf{q},M, \eta,s}  \gamma_{\mbf{k},N, \eta,s}  - \frac{1}{2} \delta_{MN} \delta_{\mbf{q},0} )
 \eea
 where the form factors $M(\mbf{k},\mbf{q})$ are defined in \Eq{eq:formfactorMTBG}. In this Appendix, we find an exact expression for the charge excitations above the groundstate defined in \Eq{eq:charge1} of the Main Text by
  \bea
\null [H_{int} - \mu N, \gamma^\dag_{\mbf{k},M, s, \eta}] \ket{\Psi_\nu} &\equiv \frac{1}{2} \sum_N  \gamma^\dag_{\mbf{k},N, s, \eta} [R^\eta_+(\mbf{k})]_{NM} \ket{\Psi_\nu}  \\
\eea
where $-\mu N$ is the chemical potential term obeying $[N, \gamma^\dag_{\mbf{k},M, s, \eta}] = \gamma^\dag_{\mbf{k},M, s, \eta}$. To compute the $[H_{int} ,\gamma^\dag_{\mbf{k},M, s, \eta}]$ we first need the commutators
\bea
\label{eq:comgamma}
\null [ O_{\mbf{q},\mbf{G}}, \gamma^\dag_{\mbf{k},M, s, \eta}]
&= \sqrt{V(\mbf{q}+2\pi \mbf{G})} \sum_N \gamma^\dag_{\mbf{k}- \mbf{q},N, \eta,s} \bar M^\eta_{N M}(\mbf{k},\mbf{q}+2\pi \mbf{G})  \\
\null [ O_{\mbf{q},\mbf{G}}, \gamma_{\mbf{k},M, s, \eta}] &=  -[O_{-\mbf{q},-\mbf{G}}, \gamma^\dag_{\mbf{k},M, s, \eta}]^\dag
= - \sqrt{V(\mbf{q}+2\pi \mbf{G})} \sum_N  \gamma_{\mbf{k} + \mbf{q},N, \eta,s} \bar {M^\eta}^*_{NM}(\mbf{k},-\mbf{q}-2\pi \mbf{G}) \ .  \\
\eea
 We will focus on the charge $+1$ excitations which arises from the $\gamma^\dag_{\mbf{k}, M, s, \eta}$ commutator. Analogous formulae for the $-1$ excitations can be obtained from the $\gamma_{\mbf{k}, M, s, \eta}$. Using \Eq{eq:comgamma}, we calculate
\bea
\null [O_{-\mbf{q},-\mbf{G}} O_{\mbf{q},\mbf{G}}, \gamma^\dag_{\mbf{k},M, s, \eta}] &=  O_{-\mbf{q},-\mbf{G}} [O_{\mbf{q},\mbf{G}}, \gamma^\dag_{\mbf{k},M, s, \eta}] + [O_{-\mbf{q},-\mbf{G}}, \gamma^\dag_{\mbf{k},M, s, \eta}] O_{\mbf{q},\mbf{G}} \\
&=   \sqrt{V(\mbf{q}+2\pi \mbf{G})} \sum_N O_{-\mbf{q},-\mbf{G}}  \gamma^\dag_{\mbf{k}- \mbf{q},N, \eta,s}  \bar{M}^\eta_{NM}(\mbf{k},\mbf{q}+2\pi \mbf{G}) \\
&\qquad +  \sqrt{V(\mbf{q}+2\pi \mbf{G})} \sum_N  \gamma^\dag_{\mbf{k}+\mbf{q},N, \eta,s} \bar{M}^\eta_{NM}(\mbf{k},-\mbf{q}-2\pi \mbf{G}) O_{\mbf{q},\mbf{G}} \\
&=  \sqrt{V(\mbf{q}+2\pi \mbf{G})} \sum_N (\gamma^\dag_{\mbf{k}- \mbf{q},N, \eta,s}  O_{-\mbf{q},-\mbf{G}} -[\gamma^\dag_{\mbf{k}- \mbf{q},N, \eta,s},  O_{-\mbf{q},-\mbf{G}}]) \bar M^\eta_{NM}(\mbf{k},\mbf{q}+2\pi \mbf{G}) \\
&\qquad +  \sqrt{V(\mbf{q}+2\pi\mbf{G})} \sum_N  \gamma^\dag_{\mbf{k}+\mbf{q},N, \eta,s} \bar M^\eta_{NM}(\mbf{k},-\mbf{q}-2\pi \mbf{G}) O_{\mbf{q},\mbf{G}} \\
\eea
Evaluating the remaining commutator, we find
\bea
\label{eq:HFterms}
\null [O_{-\mbf{q},-\mbf{G}} O_{\mbf{q},\mbf{G}}, \gamma^\dag_{\mbf{k},M, s, \eta}]  &= V(\mbf{q}+2\pi\mbf{G}) \sum_N  \gamma^\dag_{\mbf{k},N, \eta,s} [ \bar M^\eta(\mbf{k}-\mbf{q}-2\pi\mbf{G},-\mbf{q}-2\pi\mbf{G}) \bar M^\eta(\mbf{k},\mbf{q}+2\pi \mbf{G})]_{NM} \\
& \qquad +  \sqrt{V(\mbf{q}+2\pi \mbf{G})} \sum_N \lp  \gamma^\dag_{\mbf{k}+\mbf{q},N, \eta,s} \bar M^\eta_{NM}(\mbf{k},-\mbf{q}-2\pi \mbf{G}) O_{\mbf{q},\mbf{G}} + \gamma^\dag_{\mbf{k}- \mbf{q},N, \eta,s}  \bar M^\eta_{NM}(\mbf{k},\mbf{q}+2\pi \mbf{G}) O_{-\mbf{q},-\mbf{G}}\rp \ . \\
\eea
Let us focus on the form factor product $ \bar M(\mbf{k}-\mbf{q},-\mbf{q}) \bar M(\mbf{k},\mbf{q})$. Using \Eq{eq:formfactorMTBG}, we compute (suppressing the $\eta$ index temporarily)
\bea
\label{eq:PMM}
\bar M(\mbf{k}-\mbf{q},-\mbf{q})  \bar M(\mbf{k},\mbf{q})
&= e^{i \xi_\mbf{q}(\mbf{k}) +i \xi_{-\mbf{q}}(\mbf{k}-\mbf{q})}  U^\dag(\mbf{k}) \mathcal{H}^{-\mbf{q}} U(\mbf{k}-\mbf{q})  U^\dag(\mbf{k}-\mbf{q}) \mathcal{H}^{\mbf{q}} U(\mbf{k})\\
&= U^\dag(\mbf{k}) \mathcal{H}^{-\mbf{q}} U(\mbf{k}-\mbf{q}) U^\dag(\mbf{k}-\mbf{q}) \mathcal{H}^{\mbf{q}} U(\mbf{k})  \\
&= ( U^\dag(\mbf{k}-\mbf{q}) \mathcal{H}^{\mbf{q}} U(\mbf{k})   )^\dag  \, U^\dag(\mbf{k}-\mbf{q}) \mathcal{H}^{\mbf{q}} U(\mbf{k})\\
& = M^\dag(\mbf{k},\mbf{q}) M(\mbf{k},\mbf{q}) \equiv P(\mbf{k}, \mbf{q})
\eea
where we used the identity  $e^{-i \xi_\mbf{q}(\mbf{k}) - i \xi_{-\mbf{q}}(\mbf{k}-\mbf{q})} = 1$ proven in \Ref{secondpaper} and $\mathcal{H}^{-\mbf{q}} = {\mathcal{H}^{\mbf{q}} }^\dag$.  As a result of \Eq{eq:PMM},
\bea
P^\eta(\mbf{k}, \mbf{q}+2\pi \mbf{G}) \equiv {M^\eta}^\dag(\mbf{k},\mbf{q}+2\pi \mbf{G}) M^\eta(\mbf{k},\mbf{q}+2\pi \mbf{G})
\eea
 is positive semi-definite. The second line of \Eq{eq:HFterms} simplifies considerably when acting on $\ket{\Psi_\nu}$ (defined in \Eq{eq:exacteigstate} of the Main Text): 
 \bea
\label{eq:Metadisc}
 O_{\mbf{q},\mbf{G}} \ket{\Psi_\nu}  &= \delta_{\mbf{q},0} \sqrt{V(2\pi\mbf{G})} \sum_{\mbf{k}\in BZ} \sum_{\eta, s} \sum_{MN} \bar{M}^\eta_{MN}(\mbf{k},2\pi\mbf{G}) (\gamma^\dag_{\mbf{k},M, \eta,s}  \gamma_{\mbf{k},N, \eta,s}  - \frac{1}{2} \delta_{MN}) \ket{\Psi_\nu} \\
 &= \delta_{\mbf{q},0} \sqrt{V(2\pi\mbf{G})} \sum_{\mbf{k}\in BZ} \sum_{\eta, s}  (\sum_j \delta_{s,s_j} \delta_{\eta,\eta_j} \Tr  [\bar{M}^\eta(\mbf{k},2\pi\mbf{G})]  - \frac{1}{2}  \Tr  [\bar{M}^\eta(\mbf{k},2\pi\mbf{G})]) \ket{\Psi_\nu} \\
 &= \delta_{\mbf{q},0} \sqrt{V(2\pi\mbf{G})} \sum_{\mbf{k}\in BZ} (\frac{\nu+4}{2} \Tr  [\bar{M}^\eta(\mbf{k},2\pi\mbf{G})]  - \frac{4}{2} \Tr  [\bar{M}^\eta(\mbf{k},2\pi\mbf{G})] ) \ket{\Psi_\nu} \\
 &= \nu \delta_{\mbf{q},0}  \sqrt{V(2\pi\mbf{G})} \sum_{\mbf{k}\in BZ} \frac{1}{2} \Tr \bar{M}^\eta(\mbf{k},2\pi\mbf{G}) \ket{\Psi_\nu} \ .  \\
 \eea
 Returning to the second line of \Eq{eq:HFterms} and using \Eq{eq:Metadisc}, we obtain:
\bea
&  \sqrt{V(\mbf{q} + 2\pi \mbf{G})} \sum_N \lp \gamma^\dag_{\mbf{k}+\mbf{q},N, \eta,s} \bar M_{NM}(\mbf{k},-\mbf{q}- 2\pi \mbf{G})  O_{\mbf{q},\mbf{G}} + \gamma^\dag_{\mbf{k}- \mbf{q},N, \eta,s}  \bar M_{NM}(\mbf{k},\mbf{q}+ 2\pi \mbf{G})  O_{-\mbf{q},-\mbf{G}}\rp \ket{\Psi_\nu} \\
& = \nu  \delta_{\mbf{q},0}  V(2\pi\mbf{G}) \lp  \sum_{\mbf{q}'\in BZ} \frac{1}{2} \Tr \bar{M}(\mbf{q},2\pi\mbf{G}) \rp  \sum_N \gamma^\dag_{\mbf{k},N, \eta,s} \lp \bar M_{NM}(\mbf{k},-2\pi \mbf{G}) + \bar M_{NM}(\mbf{k},2\pi \mbf{G})\rp  \ket{\Psi_\nu} \ .
\eea
The sum $[\bar M(\mbf{k},-2\pi \mbf{G}) + \bar M(\mbf{k},2\pi \mbf{G})]_{NM}$ is Hermitian, which it must be because this term is part of the effective Hamiltonian. Gathering results, we find the effective Hamiltonian $\frac{1}{2}R^\eta_{\pm}(\mbf{k}) $ where $R^\eta_+(\mbf{k}) = R^\eta_F(\mbf{k}) + \nu R^\eta_H(\mbf{k}) - 2\mu \mathbb{1} $ and $R^\eta_-(\mbf{k})^* = R^\eta_F(\mbf{k}) - \nu R^\eta_H(\mbf{k}) + 2\mu \mathbb{1} $ and the various terms are defined by
\bea
\label{eq:excitationHamiltonian}
 R^\eta_F(\mbf{k})&= \sum_{\mbf{G}} \lp \frac{1}{N_M} \sum_{\mbf{q} \in BZ} \Omega^{-1}V(\mbf{q}+2\pi \mbf{G}) {M^\eta}^\dag(\mbf{k},\mbf{q}+2\pi \mbf{G}) M^\eta(\mbf{k},\mbf{q}+2\pi \mbf{G}) \rp , \\
R^\eta_H(\mbf{k}) &=  \sum_{\mbf{G}\neq 0} \Omega^{-1}V(2\pi \mbf{G}) \lp \frac{1}{N_M}  \sum_{\mbf{q}\in BZ} \frac{1}{2} \Tr M^\eta(\mbf{q},2\pi\mbf{G}) \rp  M^\eta(\mbf{k},2\pi \mbf{G}) + H.c. ,\\
\mu &= \nu \sum_{\mbf{G}\neq 0} \Omega^{-1}V(2\pi \mbf{G}) \lp \frac{1}{N_M}  \sum_{\mbf{q}\in BZ} \frac{1}{2} \Tr M^\eta(\mbf{q},2\pi\mbf{G}) \rp^2 \
\eea
where we have discretized the Brillouin zone by taking $N_M$ to be the (finite) number of moir\'e unit cells, so each $\mbf{q}$ sum has $N_M$ terms. The value of the chemical potential is derived using the flat metric approximation in \Ref{secondpaper}. Importantly, the Fock term $R_F(\mbf{k})$ is positive definite \cite{2020arXiv200914200B}. We conclude that the groundstate $\ket{\Psi_0}$ where $\nu =0$ is stable to charge excitations and generically (but not always) will be insulating. Lastly, we note that
\bea
R^{-\eta}_F(\mbf{k}) &= \nu_1  R^{\eta}_F(\mbf{k}) \nu_1, \qquad R^{-\eta}_H(\mbf{k}) = \nu_1 R^{\eta}_H(\mbf{k}) \nu_1 \\
\eea
where $\nu_1$ is a Pauli matrix, so charge $+1$ and charge $-1$ spectra are related by the unitary matrix $\nu_1$ and hence have identical spectra.

\begin{figure*}[h]
\centering
\includegraphics[height = 4.2cm]{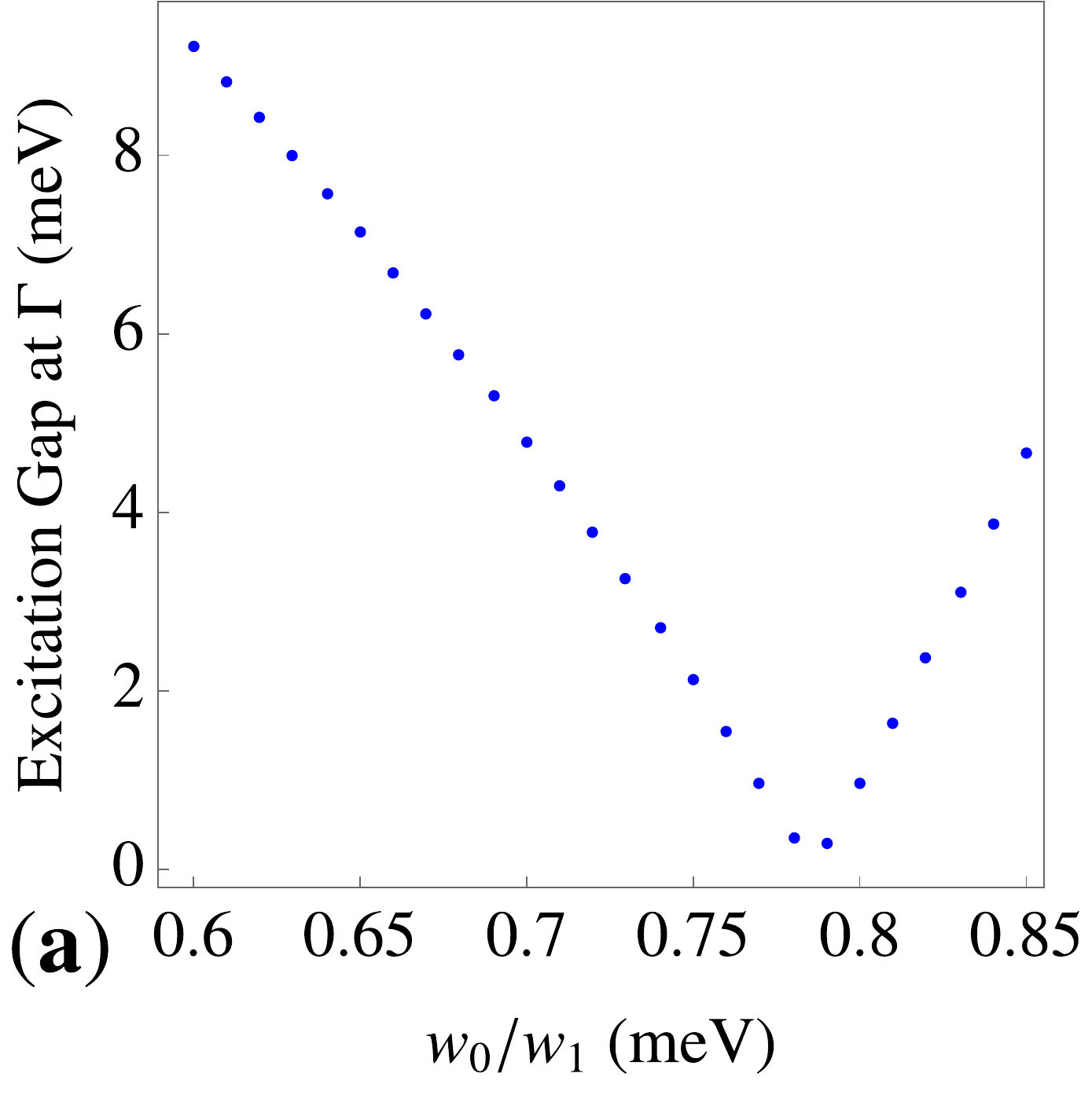}
\includegraphics[height = 4.2cm]{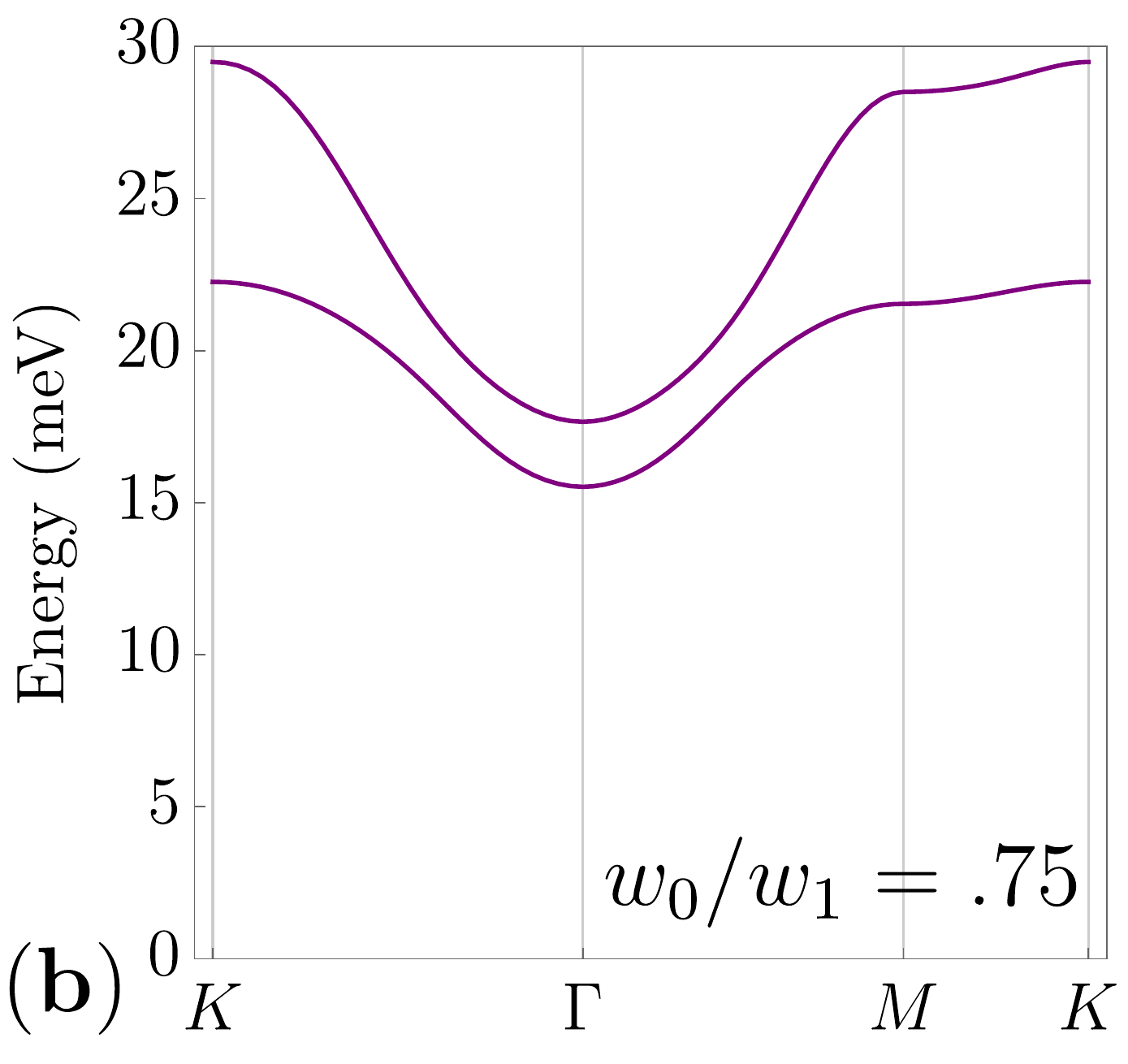}
\includegraphics[height = 4.2cm]{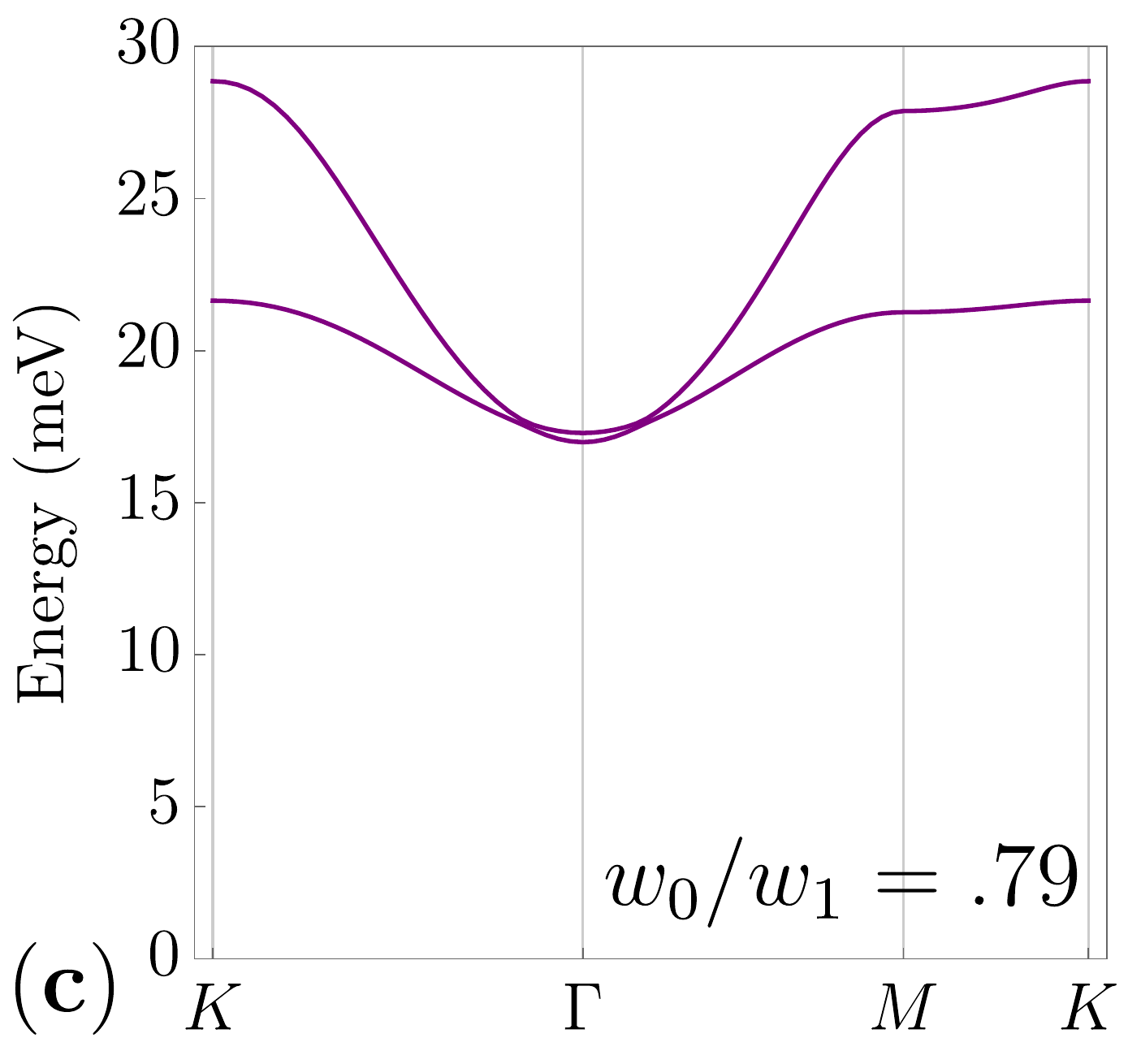}
\includegraphics[height = 4.2cm]{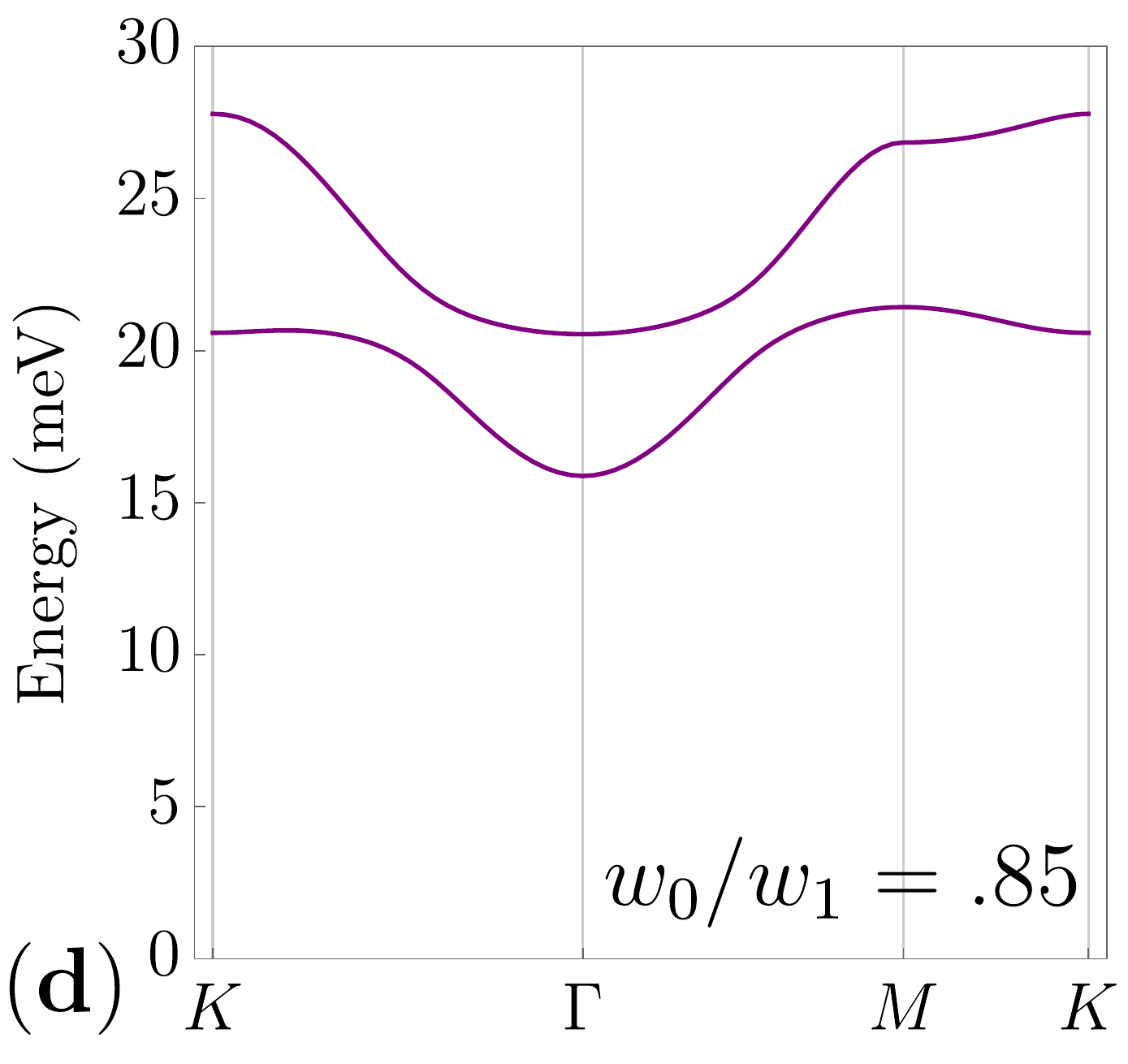}
\caption{Charge $\pm1$ excitations at $\nu=0$ using $\xi = 10$nm in \Eq{eq:paramapp}. $(a)$ The excitation gap of $R^\eta_{\pm}(\mbf{k}=0)$ is plotted as a function of $w_0$, showing that the near-degeneracy of the bands at $w_0/w_1 = .8$ is accidental. $(b)-(d)$ display the band structure at $w_0/w_1 = .75, .79, .85$ respectively, highlighting the gap closing and reopening at $\Gamma$ as $w_0/w_1$ is increased. }
\label{fig:excit}
\end{figure*}

In \Fig{fig:excit}, we study the dispersion relation of the quasi-particles near $\nu=0$ obtained from the charge $\pm1$ excitation Hamiltonian $R^\eta_{\pm}(\mbf{k})$. We find that the nearly degenerate bands at the $\Gamma$ point for $w_0/w_1 = .8$ (shown in Fig. 2 of the Main Text) are not generic, which is understood from the symmetries. At zero flux, \Ref{2020arXiv200914200B} found that the excitation bands had a protected degeneracy at the $\Gamma$ point due to $C_{2z}\mathcal{T}$ symmetry, but nonzero flux breaks this symmetry and allows the bands to gap. \Fig{fig:excit}a shows the gap between the two excitation bands at $\Gamma$ as a function of $w_0$, from which we see that $w_0 = .8 w_1$ is in a non-generic region in the parameter space where the bands are close in energy. \Fig{fig:excit}b-d show three examples of excitation band structures.

\end{document}